# HySemRAG: A Hybrid Semantic Retrieval-Augmented Generation Framework for Automated Literature Synthesis and Methodological Gap Analysis

A Preprint

**Alejandro Godinez**
Department of Environmental Health Sciences
University at Albany, State University of New York
Albany, 12222

Bandit Environmental Solutions LLC
Albany
agodinez@albany.edu

2025-07-31

## Abstract

We present HySemRAG, a framework that combines Extract, Transform, Load (ETL) pipelines with Retrieval-Augmented Generation (RAG) to automate large-scale literature synthesis and identify methodological research gaps. The system addresses limitations in existing RAG architectures through a multi-layered approach: hybrid retrieval combining semantic search, keyword filtering, and knowledge graph traversal; an agentic self-correction framework with iterative quality assurance; and post-hoc citation verification ensuring complete traceability. Our implementation processes scholarly literature through eight integrated stages: multi-source metadata acquisition, asynchronous PDF retrieval, custom document layout analysis using modified Docling architecture, bibliographic management, LLM-based field extraction, topic modeling, semantic unification, and knowledge graph construction. The system creates dual data products - a Neo4j knowledge graph enabling complex relationship queries and Qdrant vector collections supporting semantic search - serving as foundational infrastructure for verifiable information synthesis. Evaluation across 643 observations from 60 testing sessions demonstrates structured field extraction achieving 35.1% higher semantic similarity scores ($0.655 \pm 0.178$) compared to PDF chunking approaches ($0.485 \pm 0.204$, $p < 0.000001$). The agentic quality assurance mechanism achieves 68.3% single-pass success rates with 99.0% citation accuracy in validated responses. Applied to geospatial epidemiology literature on ozone exposure and cardiovascular disease, the system identifies methodological trends and research gaps, demonstrating broad applicability across scientific domains for accelerating evidence synthesis and discovery.

## 1 Introduction

Scientific literature synthesis has become challenging as publication rates accelerate and research domains become more interdisciplinary. Traditional systematic review processes are time-consuming and labor-intensive, often requiring months to complete comprehensive analyses [Gue et al., 2024, Tsai et al., 2024]. Retrieval-Augmented Generation (RAG) systems offer solutions by combining parametric memory from large language models with non-parametric external knowledge bases [Li and Lai, 2024]. However, existing RAG implementations face limitations including noisy retrieval, hallucination in generated responses, and lack of verifiable citations [Barnett et al., 2024, Huang et al., 2024].

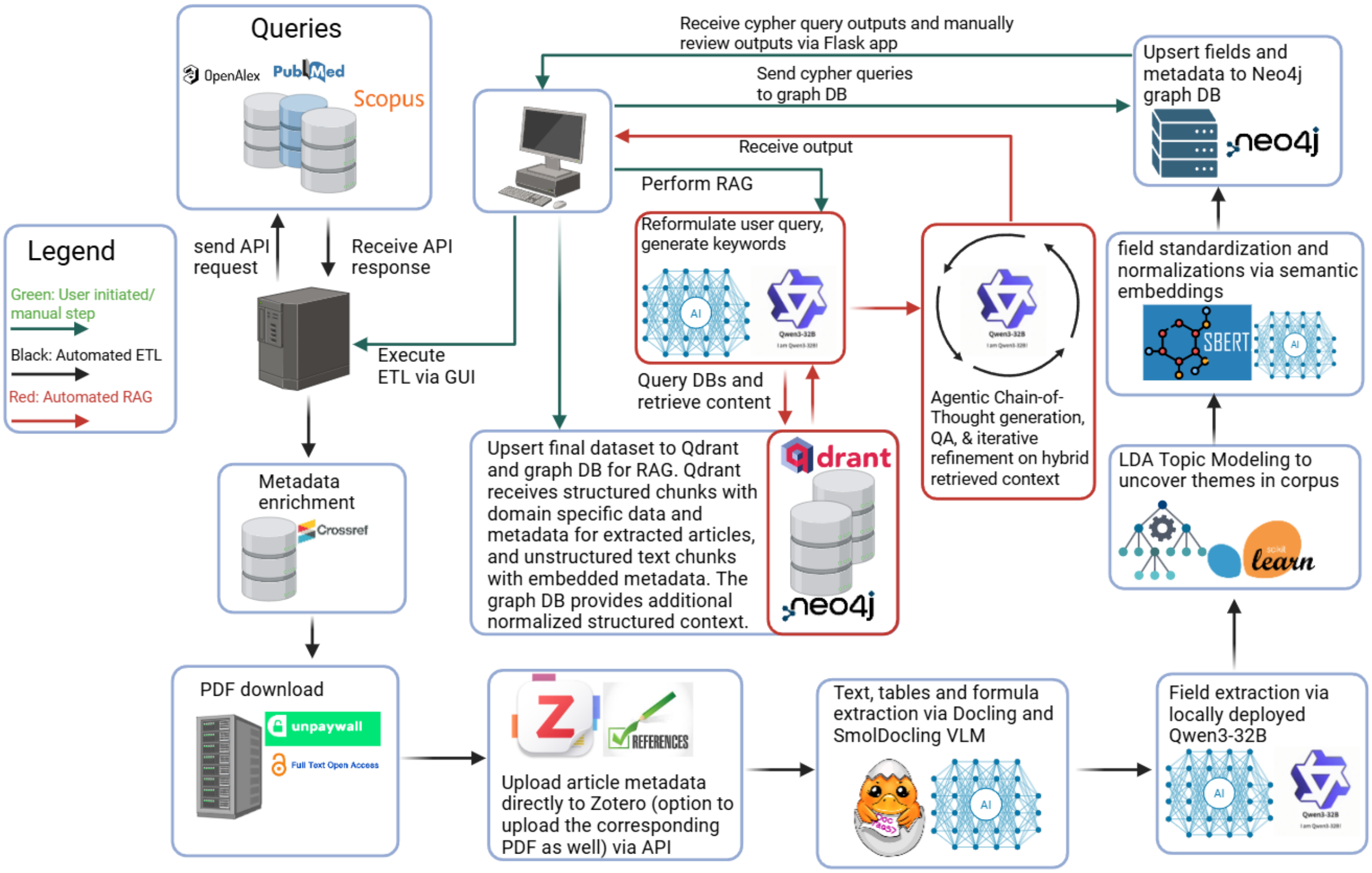

Figure 1: Graphical Abstract

The identification of methodological gaps represents a challenging application for automated literature analysis. Unlike simple information retrieval, gap analysis requires understanding research methodology distributions, identifying underexplored approaches, and synthesizing evidence across multiple dimensions [Barnett et al., 2024]. Current approaches often rely on manual coding or simple keyword analysis, limiting their scope and introducing human bias [J. Bolaños et al., 2024].

We address these challenges through HySemRAG, a framework that integrates ETL processing with multi-agent RAG capabilities. Our approach moves beyond traditional single-pass generation by implementing: (1) hybrid retrieval strategies that combine semantic search, structured queries, and knowledge graph traversal; (2) iterative quality assurance through multi-agent validation loops; and (3) comprehensive citation verification ensuring complete provenance tracking.

The framework's design enables systematic identification of methodological gaps by constructing detailed knowledge graphs from literature corpora and supporting complex analytical queries across research domains. We demonstrate the system's capabilities through application to geospatial epidemiology literature examining ozone exposure and cardiovascular disease, revealing specific methodological trends and underexplored research directions.

## 2 Related Work

### 2.1 Retrieval-Augmented Generation in Scientific Literature

RAG frameworks combine deep learning with traditional information retrieval, integrating parametric and non-parametric memory systems to enhance text generation accuracy [Feng et al., 2024]. In scientific applications, RAG systems have demonstrated efficiency gains, with Gu et al. [2024] reporting reduction in systematic review extraction time from 1,310 minutes to 5 minutes while maintaining data quality. Similarly, Kreimeyer et al. [2024] achieved over 80% reproduction of expert-curated information in oncology literature.

Performance evaluations reveal RAG's superiority over conventional approaches. Fateh Ali et al. [2024] demonstrated RAG achieving higher ROUGE-1 (0.364) and ROUGE-2 (0.123) scores compared to transformer-only models and frequency-based methods. Upadhyay and Viviani [2025] showed RAG nearly doubling performance metrics like CAM NDCG (0.2146 vs. 0.1119) in health information retrieval tasks.

### 2.2 Methodological Gap Analysis and Evidence Synthesis

Traditional gap analysis relies heavily on manual systematic reviews and meta-analyses [Fleurence et al., 2024]. Recent developments in AI-driven approaches have begun addressing scalability limitations. Barnett et al. [2024] demonstrated RAG's capability as a meta-analytic tool for identifying methodological weaknesses across research workflows. Maharana et al. [2025] showed how RAG analysis can reveal under-explored techniques like quantized LLMs offering efficiency benefits overlooked in conventional approaches.

Knowledge graph-based methods promote interdisciplinary collaboration by mapping research concepts across domains [J. Bolaños et al., 2024]. Di Maria et al. [2024] created interactive knowledge graphs from biomedical literature enabling shared exploration by diverse specialists. Dernbach et al. [2024] demonstrated improved multi-hop reasoning across distinct corpora through LLM-knowledge graph alignment.

### 2.3 Limitations of Current Approaches

Existing RAG implementations face several challenges. Barnett et al. [2024] identified seven failure points including retrieval failures, context limitations, and hallucination issues. Huang et al. [2024] emphasized factual inaccuracies as threats to scientific rigor. Citation accuracy remains problematic, with many systems generating unreliable references [Upadhyay and Viviani, 2025].

Current gap analysis approaches often lack systematic frameworks for identifying methodological trends across large literature corpora. Manual approaches introduce human bias and scalability constraints [Tsai et al., 2024], while simple automated methods miss nuanced methodological patterns requiring deep domain understanding.

## 3 Methods

### 3.1 System Architecture Overview

HySemRAG implements an eight-stage ETL pipeline integrated with a multi-agent RAG framework. The system transforms raw scholarly literature into structured, queryable knowledge representations through: (1) multi-source data acquisition, (2) asynchronous PDF retrieval, (3) custom document layout analysis, (4) bibliographic management, (5) LLM-based field extraction, (6) topic modeling, (7) semantic unification, and (8) knowledge graph construction and vector database indexing.

### 3.2 Corpus Size and Variability

The final corpus size in each ETL run varies based on the availability of open-access full-text PDFs via the Unpaywall API. From an initial candidate pool of approximately 3,400 metadata records retrieved from PubMed, OpenAlex, and Scopus, the number of successfully retrieved full-text documents typically ranges from 500 to 1,300. This variation reflects the constraints of automated scholarly literature retrieval that depends on open-access publications. The processing stages maintain consistent performance (accuracy of data extraction, knowledge graph quality, and output consistency) regardless of corpus size. This makes the ETL pipeline useful for research scenarios where subscription-based access is limited.

**Demonstration Video**

Here is a video demonstrating the system:

https://youtu.be/ZCy5ESJ1gVE?si=K8CttwgTj7yGrWjn

### 3.3 Stage 1: Multi-Source Data Acquisition and Enrichment

The first stage of the ETL pipeline acquires scholarly literature from multiple sources. This phase gathers article metadata from different databases, each with unique strengths, and combines them into a single, deduplicated dataset. Python modules handle this process to ensure the subsequent pipeline stages operate on clean data.

#### 3.3.1 Parallelized Metadata Fetching

The system queries three scholarly databases: PubMed for biomedical literature, OpenAlex for multidisciplinary coverage and citation data, and Scopus for abstracts and citations in STEM fields. The GUI allows users to select any combination of these sources.

Since sequential API calls for thousands of articles would create a performance bottleneck, the system uses concurrent data retrieval. The implementation handles the specific rate limits of each API. For PubMed, the scripts/fast_pubmed.py module performs an esearch call to retrieve up to 10,000 PMIDs, then uses ThreadPoolExecutor to fetch article metadata in parallel batches via efetch while staying within NCBI's 3 queries per second limit. For OpenAlex and Scopus, the scripts/fast_openalex.py and scripts/etl_elsevier.py modules use asyncio and aiohttp for asynchronous requests. This allows hundreds of concurrent HTTP requests without blocking on network latency. An asyncio.Semaphore throttles requests to stay within each API's rate limits.

#### 3.3.2 Data Fusion and Deduplication:

After parallel fetching completes, the metadata from each source is held in separate pandas DataFrames and must be combined into a single dataset. The merge_and_deduplicate function in scripts/etl_utils_refined.py handles this by concatenating the DataFrames into one dataset.

```
#| label: merge-deduplicate
#| eval: false

def merge_and_deduplicate(
    from_pubmed_all: List[pd.DataFrame],
    from_openalex_all: List[pd.DataFrame],
    from_elsevier_all: Optional[List[pd.DataFrame]] = None
) -> pd.DataFrame:
    all_dfs = []
    if from_pubmed_all:
        all_dfs.extend(from_pubmed_all)
    if from_openalex_all:
        all_dfs.extend(from_openalex_all)
    if from_elsevier_all:
        all_dfs.extend(from_elsevier_all)
    if not all_dfs:
        logging.warning("No DataFrames provided to merge_and_deduplicate.")
        return pd.DataFrame()
    combined_df = pd.concat(all_dfs, ignore_index=True)
    if combined_df.empty:
        logging.info("Combined DataFrame is empty before deduplication.")
        return combined_df
    deduplicated_df = deduplicate_by_doi_title(combined_df)
    return deduplicated_df
```

The next step is deduplication, since the same article may appear in multiple databases. The system uses a two-pass deduplication algorithm. First, it normalizes all DOIs by converting to lowercase and removing whitespace, then retains the first occurrence of each unique DOI.

```
def deduplicate_by_doi_title(df: pd.DataFrame) -> pd.DataFrame:
    # Create normalized columns
    if "DOI" not in df.columns: df["DOI"] = ""
    if "Title" not in df.columns: df["Title"] = ""
    df["DOI"] = df["DOI"].fillna("")
    df["Title"] = df["Title"].fillna("")

    df["DOI_clean"] = df["DOI"].astype(str).str.lower().str.strip()
    df["Title_clean"] = df["Title"].astype(str).str.lower().str.strip()
    df.sort_values(by=["DOI_clean", "Title_clean"], inplace=True, kind='stable')
    indices_to_keep = set()
    seen_dois = set()
    for index, doi in df.loc[df["DOI_clean"] != "", "DOI_clean"].items():
        if doi not in seen_dois:
            indices_to_keep.add(index)
            seen_dois.add(doi)
    empty_doi_mask = (df["DOI_clean"] == "") & (~df.index.isin(indices_to_keep))
    seen_titles_empty_doi = set()
    for index, title in df.loc[empty_doi_mask, "Title_clean"].items():
        if title not in seen_titles_empty_doi:
            indices_to_keep.add(index)
            seen_titles_empty_doi.add(title)
    df_deduplicated = df.loc[list(indices_to_keep)].copy()
    df_deduplicated.drop(columns=["DOI_clean", "Title_clean"], inplace=True)
    df_deduplicated.reset_index(drop=True, inplace=True)
    return df_deduplicated
```

For records without DOIs, the algorithm performs a second pass using normalized article titles. This ensures each scholarly work appears only once in the final dataset.

#### 3.3.3 Metadata Enrichment:

The merged dataset may contain incomplete records (for example, a PubMed record might lack a PDF link that OpenAlex provides). The scripts/metadata_enrichment.py module addresses this by taking all unique DOIs from the deduplicated dataset and querying both OpenAlex and Crossref APIs using an asynchronous approach.

```
#| label: enrich-metadata
#| eval: false

def enrich_metadata_sync(
    df: pd.DataFrame,
    crossref_email: Optional[str] = None
) -> pd.DataFrame:
    email = crossref_email or MAILTO
    dois = df["DOI"].dropna().unique().tolist()
    enrichment = asyncio.run(enrich_metadata(dois, email))

    for idx, row in df.iterrows():
        doi = row["DOI"]
        metadata = enrichment.get(doi, {})
        if metadata:
            if pd.isna(row.get("Abstract")) or not str(row["Abstract"]).strip():
                df.at[idx, "Abstract"] = metadata.get("Abstract", row["Abstract"])

            if (pd.isna(row.get("ItemType")) or
                not str(row["ItemType"]).strip() or
                row["ItemType"] == "N/A"):
                df.at[idx, "ItemType"] = metadata.get("ItemType",
                                                      row.get("ItemType", "N/A"))

            citation_count = metadata.get("CitationCount", row.get("CitationCount"))
            df.at[idx, "CitationCount"] = (int(citation_count)
                                           if citation_count is not None else 0)

            primary_topic = metadata.get("primary_topic")
            if primary_topic:
                df.at[idx, "primary_topic"] = json.dumps(primary_topic)
            else:
                df.at[idx, "primary_topic"] = pd.NA

            df.at[idx, "is_published"] = metadata.get("is_published",
                                                      row.get("is_published"))
            df.at[idx, "is_retracted"] = metadata.get("is_retracted",
                                                      row.get("is_retracted"))
            df.at[idx, "OpenAlexID"] = metadata.get("OpenAlexID",
                                                    row.get("OpenAlexID"))

    df["CitationCount"] = df["CitationCount"].fillna(0).astype(int)
    return df
```

The results are merged back into the main DataFrame without overwriting existing data. For example, an abstract from Crossref is added only if the record's Abstract field is empty. Fields like ItemType or CitationCount are updated only if the existing data is null or a placeholder. This produces the most complete version of each record by combining data from all sources.

### 3.4 Stage 2: Asynchronous Full-Text PDF Retrieval

After creating a deduplicated list of scholarly articles, the next stage acquires full-text content for each entry. This is needed for the content analysis, entity extraction, and topic modeling in later stages. The system uses each article's DOI to query the Unpaywall API, which indexes legally available open-access publications. Since this involves thousands of network requests, the system uses an asynchronous architecture to handle this I/O-bound task.

The scripts/async_unpaywall.py module uses Python's asyncio and aiohttp libraries. This allows the system to manage thousands of concurrent HTTP requests without being blocked by network latency. To comply with Unpaywall's API usage policies, an asyncio.Semaphore limits the request rate. The semaphore uses the API_RATE_LIMIT value from the configuration file (e.g., 8 queries per second). Each worker task must acquire the semaphore before making an API call, ensuring the request rate stays within the limit. The worker function queries Unpaywall for each DOI to get potential PDF locations, prioritizes the best_oa_location, and attempts to download from each URL until a valid PDF is found.

```python
# In scripts/async_unpaywall.py

async def _worker(row_idx, doi, title, session):
    """Return (row_idx, pdf_path|None, status_string)."""
    api_url = f"https://api.unpaywall.org/v2/{doi}?email={UNPAYWALL_EMAIL}"
    # _fetch_json is rate-limited by a semaphore
    data = await _fetch_json(session, api_url)
    if not data:
        return row_idx, None, "Unpaywall_fail"

    # Build a list of candidate URLs, prioritizing the best_oa_location
    candidates = []
    if (b := data.get("best_oa_location")): candidates.append(b)
    candidates += data.get("oa_locations", [])

    for loc in candidates:
        pdf_url = loc.get("url_for_pdf")
        if not pdf_url:
            continue
        # Attempt to download the PDF from the candidate URL
        pdf_bytes = await _download_pdf(session, pdf_url)
        if pdf_bytes:
            # Sanitize filename and save the file
            safe_name = re.sub(r'[\\/*?:"<>|()]+', '', title)[:100] or doi
            fname = f"{safe_name}_{row_idx}.pdf"
            path  = os.path.join(PDF_SAVE_FOLDER, fname)
            with open(path, "wb") as f:
                f.write(pdf_bytes)
            return row_idx, path, "Saved"

    return row_idx, None, "No_PDF"
```

When a PDF downloads successfully, it's saved to a local directory specified in the configuration, and the DataFrame is updated with the local file path for that record. Records without retrievable full-text PDFs are filtered out at the end of this stage. The resulting DataFrame contains only articles with local PDF files and is passed to the next stage for content extraction.

### 3.5 Stage 3: Bibliographic Management and Citation Rendering

After acquiring article metadata and retrieving full-text PDFs, the pipeline creates a centralized library using the Zotero reference management system and generates citation metadata for each article. This citation data provides verification for the downstream Retrieval-Augmented Generation (RAG) agent. The scripts/fast_zotero_gui.py module handles this process.

The Zotero integration processes articles in batches. The integrate function receives the DataFrame of articles with valid PDF paths and uses the pyzotero library to communicate with the Zotero API. New items are created in batches of up to 50 articles per API call. The system uses ThreadPoolExecutor to upload PDF files to the Zotero entries in parallel.

The citation generation uses the citeproc-py library to render citations locally rather than making repeated API calls to Zotero. This approach uses Citation Style Language (CSL) files (e.g., apa.csl, ieee.csl), selectable in the GUI, to format both full bibliographic citations and in-text citations for each article. The module includes a fallback mechanism that queries the Zotero API if local citeproc rendering fails for a specific entry.

```
# In scripts/fast_zotero_gui.py

# ... inside the integrate function loop ...
for idx, zkey in attach.items():
    df.at[idx, "ZoteroKey"] = zkey
    meta = {c: df.at[idx, c] for c in (
        "Title", "DOI", "Authors", "Date",
        "Journal", "Volume", "Issue", "Pages"
    )}
    try:
        # Attempt local citation rendering first
        in_text, full = _render_citations(meta, csl_path)
    except Exception as exc:
        # If local rendering fails, use the fallback
        logging.error(f"citeproc rendering failed for row {idx}: {exc}")
        in_text, full = fallback(zkey, csl_path) if callable(fallback) else ("", "")
    df.at[idx, "InTextCitation"] = in_text or ""
    df.at[idx, "FullCitation"] = full or ""
```

The generated ZoteroKey, InTextCitation, and FullCitation are added back to the DataFrame. This data is passed as context to the downstream RAG agent.

This design ensures the accuracy of the final output. The RAG agent must populate its findings into a structured JSON object that requires these metadata fields for each piece of evidence. By providing pre-generated citation data in the context, the system prevents the Large Language Model from generating its own citations, which are often unreliable.

```
// Example Metadata Schema for a single RAG Observation
{
  "metadata": {
    "PDF_DocIndex": "148",
    "PDF_ChunkIndex": "0",
    "Struct_DocIndex": "N/A",
    "Struct_ChunkIndex": "N/A",
    "DOI": "10.1016/j.jobe.2021.103722",
    "ZoteroKey": "ABC123DE",
    "InTextCitation": "(Author et al., 2021)",
    "FullCitation": "Author, A., Author, B., & Author, C. (2021). " +
                    "Title of the article. Journal Name, 1(2), 100-110."
  }
}
```

This approach prevents the LLM from generating potentially incorrect citations and ensures every piece of evidence in the final answer links back to a specific source document in the Zotero library.

### 3.6 Stage 4: Content Extraction via Document Layout Analysis

**Objective: High-Fidelity Extraction from Scholarly PDFs**

The ETL pipeline needs to extract content from thousands of scholarly articles in PDF format. This is challenging due to PDFs' complex, variable layouts and the need to preserve structural context. This stage goes beyond simple text extraction to parse the full textual body of each article plus embedded elements like tabular data and mathematical formulas in LaTeX format.

#### 3.6.1 Core Technology and Identified Limitations

The system uses IBM's Docling library as its document analysis engine, which uses machine learning for page layout detection. Initial testing revealed several limitations that required substantial re-engineering:

1. **Performance Bottleneck:** The framework lacked native support for parallelization, making processing thousands of documents slow and single-threaded.
2. **Model Inefficiency:** The original formula recognition model (SamOPTForCausalLM architecture) required approximately 30 GB of VRAM per instance, limiting parallelization.
3. **Layout Inaccuracies:** The layout detection model produced poor bounding boxes:
    - **Formula Fragmentation:** Single, multi-line mathematical formulas were incorrectly segmented into several independent clusters.
    - **Layout Misclassification:** Pages with structured elements like line numbers were often misclassified as a single, large TABLE cluster, losing semantic information.
4. **Stability Issues:** Glyph-parsing errors within the core library caused instability during large-scale processing.

To address these challenges, the Docling library was substantially re-engineered. The modified codebase, testing scripts, sample data, and validation notebook are available in a public GitHub repository.

Github Repository

#### 3.6.2 Custom Enhancements and Implementation

To overcome these challenges, the system was modified through several enhancements, handled by scripts/docling_multi_mp_gui.py.

The performance bottleneck was addressed by building a custom architecture around Python's concurrent.futures.ProcessPoolExecutor. A custom worker_initializer function enables parallel processing across multiple GPUs. This initializer uses a locking mechanism to assign each worker process to a specific GPU from the available hardware, converting Docling into a distributed application and reducing processing time.

The formula recognition model issues were solved by replacing the model entirely. The core code_formula_predictor.py was reconfigured to use "SmolDocling," which reduced VRAM requirements to ~8GB.

The layout inaccuracies and stability issues were addressed through direct modifications to the Docling codebase. This included a post-processing step in docling/utils/layout_postprocessor.py with rule-based heuristics to merge fragmented formulas and re-classify misidentified pages. The glyph errors required debugging the compiled C-extension library (pdf_parsers.cpython-312-x86_64-linux-gnu.so) using the Ghidra reverse engineering suite to diagnose and patch the root cause.

These modifications affected numerous library files:

- **Primary Docling Library Modifications:**
    - docling/backend/docling_parse_v4_backend.py
    - docling/datamodel/base_models.py
    - docling/datamodel/pipeline_options.py
    - docling/models/code_formula_model.py
    - docling/models/layout_model.py
    - docling/pipeline/vlm_pipeline.py
    - docling/utils/layout_postprocessor.py
- **IBM Model Predictor Modifications:**
    - docling_ibm_models/code_formula_model/code_formula_predictor.py
    - docling_ibm_models/layoutmodel/layout_predictor.py

### 3.6.3 Initial Challenge: Inaccurate Bounding Boxes and Model Inefficiency

The first issue was the precision of the layout analysis model. The bounding boxes generated for mathematical formulas were often too tight or included visual artifacts from surrounding text. When these imperfect image snippets were passed to the original formula recognition model (ds4sd/CodeFormula), it frequently failed, resulting in repetitive, malformed LaTeX output (as exemplified in Figure 2, showing a poorly cropped snippet).

$$\% - \text{change} = \left[ \frac{\text{IS without cover} - \text{IS with cover}}{\text{IS without cover}} \right] \times 100 \quad (17)$$

Figure 2: Bad Formula

The original model, based on the SamOPTForCausalLM architecture, required approximately 30 GB of VRAM per instance. This made large-scale parallelization across multiple GPUs expensive and inefficient.

### 3.6.4 Bounding Box and Image Preprocessing Modifications

The initial attempt to resolve the formula recognition issue focused on improving the quality of image snippets before they were sent to the model. This involved modifications to docling/models/code_formula_model.py.

- **Dynamic Bounding Box Expansion:** Logic was added to the prepare_element method to expand the bounding boxes of detected formulas. Instead of using the original, tightly-cropped box, the new implementation expands the box by a configurable ratio of the snippet's height and width. This captures more visual context as demonstrated in Figure 3.

$$\% - \text{change} = \left[ \frac{\text{IS without cover} - \text{IS with cover}}{\text{IS without cover}} \right] \times 100$$

Figure 3: Good Formula

- **Ratio-Based Image Padding:** The original _pad_with_most_frequent_edge_color function, which used static pixel padding, was replaced. The new version accepts floating-point ratios, allowing it to apply padding that is proportional to the image's dimensions. This creates more consistent input for the model, regardless of the formula's original size.
- **Conditional Image Masking:** To isolate formulas from surrounding text, conditional logic was added to use Docling's page.get_masked_image() method for items labeled as DocItemLabel.FORMULA.

#### 3.6.5 Implementation of Selective Content Masking

The most critical enhancement was the development of a method to isolate formula elements from adjacent text. When the layout model produced imperfect bounding boxes, snippets would often contain partial lines of text from the main body, confusing the formula recognition model. To solve this, a selective masking capability was added to the Page class within docling/datamodel/base_models.py.

This was achieved by introducing two new methods, get_masked_image and the internal _create_masked_image. The _create_masked_image function takes a complete, high-resolution image of a document page. It then iterates through all layout clusters identified by the layout model. For every cluster that is not labeled as a DocItemLabel.FORMULA, the function draws a white rectangle over its bounding box.

```python
def _create_masked_image(
    self,
    scale: float,
    pdf_identifier: Optional[str] = None
) -> Optional[Image]:
    try:
        original_image = self.get_image(scale=scale)
        if original_image is None or self.size is None:
            _log.warning("Original image or size is None for page %s.",
                         self.page_no)
            return None

        masked = original_image.copy()
        draw = ImageDraw.Draw(masked)
        scale_x = masked.width / self.size.width
        scale_y = masked.height / self.size.height
        top_expansion_factor = 0.045
        bottom_expansion_factor = 0.045
        cluster_count = len(self.predictions.layout.clusters)

        _log.debug("Creating masked image for page %s with %d clusters.",
                   self.page_no, cluster_count)

        # Mask non-formula clusters explicitly here with correct scaling
        for cl in self.predictions.layout.clusters:
            if cl.label != DocItemLabel.FORMULA:
                bbox = cl.bbox.to_top_left_origin(
                    page_height=self.size.height
                )
                bbox_height = bbox.b - bbox.t

                expanded_bbox = BoundingBox(
                    l=bbox.l,
                    t=bbox.t - bbox_height * top_expansion_factor,
                    r=bbox.r,
                    b=bbox.b + bbox_height * bottom_expansion_factor,
                    coord_origin=bbox.coord_origin,
                )

                scaled_bbox = BoundingBox(
                    l=expanded_bbox.l * scale_x,
                    t=expanded_bbox.t * scale_y,
                    r=expanded_bbox.r * scale_x,
                    b=expanded_bbox.b * scale_y,
                    coord_origin=expanded_bbox.coord_origin
                )

                draw.rectangle(scaled_bbox.as_tuple(), fill="white")
```

```
    return masked
```

The result is a new, "masked" image of the page where all non-formula content has been obscured, as demonstrated in the comparison between an original page layout (Figure 4) and its masked counterpart (Figure 5).

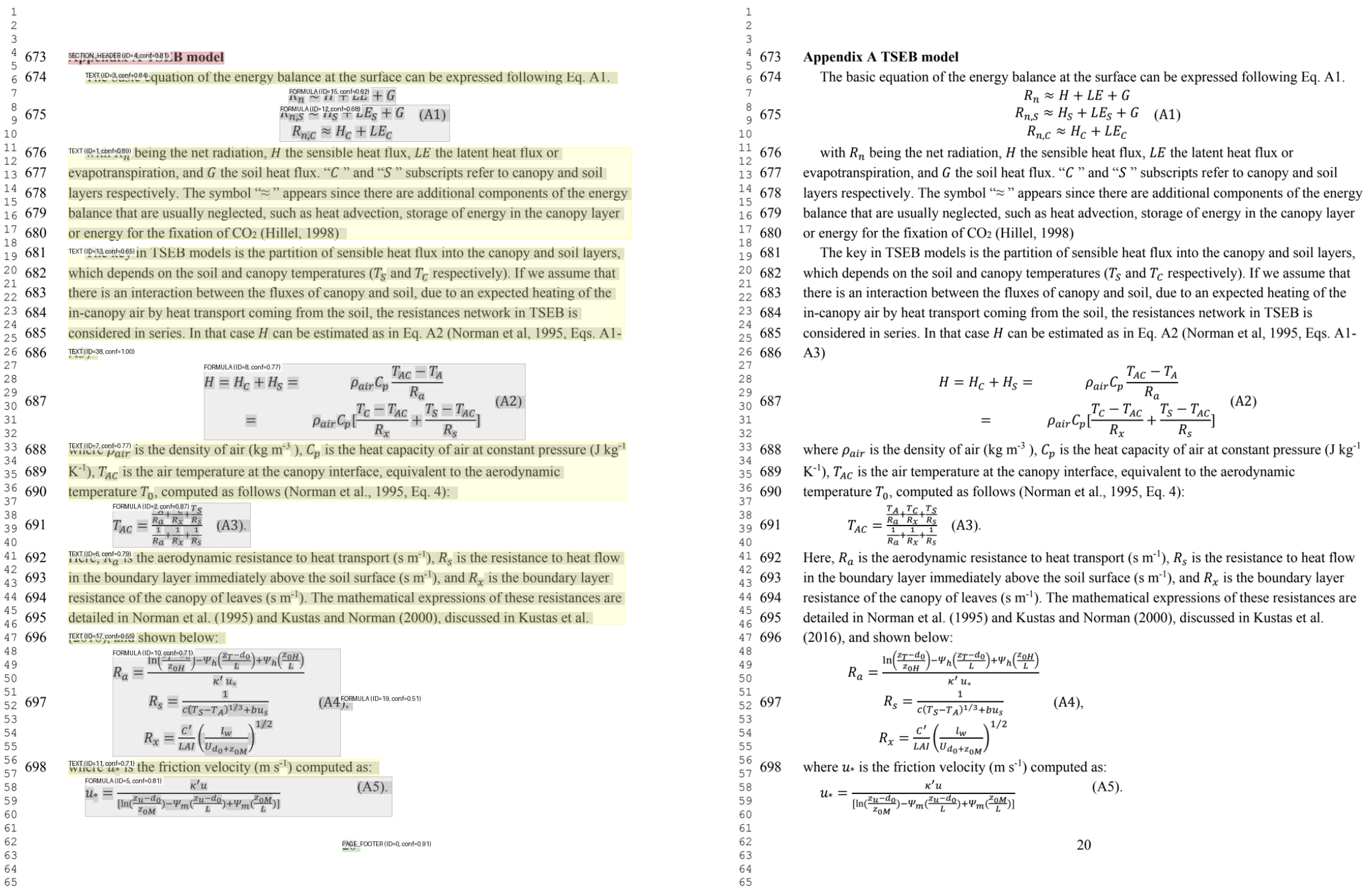

**Appendix A TSEB model**

The basic equation of the energy balance at the surface can be expressed following Eq. A1.

$$R_n \approx H + LE + G$$
$$R_{n,S} \approx H_S + LE_S + G \quad \text{(A1)}$$
$$R_{n,C} \approx H_C + LE_C$$

with $R_n$ being the net radiation, $H$ the sensible heat flux, $LE$ the latent heat flux or evapotranspiration, and $G$ the soil heat flux. "$C$ " and "$S$ " subscripts refer to canopy and soil layers respectively. The symbol "≈ " appears since there are additional components of the energy balance that are usually neglected, such as heat advection, storage of energy in the canopy layer or energy for the fixation of $CO_2$ (Hillel, 1998)

The key in TSEB models is the partition of sensible heat flux into the canopy and soil layers, which depends on the soil and canopy temperatures ($T_S$ and $T_C$ respectively). If we assume that there is an interaction between the fluxes of canopy and soil, due to an expected heating of the in-canopy air by heat transport coming from the soil, the resistances network in TSEB is considered in series. In that case $H$ can be estimated as in Eq. A2 (Norman et al, 1995, Eqs. A1-A3)

$$H = H_C + H_S = \rho_{air} C_p \frac{T_{AC} - T_A}{R_a} = \rho_{air} C_p [\frac{T_C - T_{AC}}{R_x} + \frac{T_S - T_{AC}}{R_s}] \quad \text{(A2)}$$

where $\rho_{air}$ is the density of air (kg $m^{-3}$ ), $C_p$ is the heat capacity of air at constant pressure (J $kg^{-1}$ $K^{-1}$), $T_{AC}$ is the air temperature at the canopy interface, equivalent to the aerodynamic temperature $T_0$, computed as follows (Norman et al., 1995, Eq. 4):

$$T_{AC} = \frac{\frac{T_A}{R_a}+\frac{T_C}{R_x}+\frac{T_S}{R_S}}{\frac{1}{R_a}+\frac{1}{R_x}+\frac{1}{R_S}} \quad \text{(A3).}$$

Here, $R_a$ is the aerodynamic resistance to heat transport (s $m^{-1}$), $R_s$ is the resistance to heat flow in the boundary layer immediately above the soil surface (s $m^{-1}$), and $R_x$ is the boundary layer resistance of the canopy of leaves (s $m^{-1}$). The mathematical expressions of these resistances are detailed in Norman et al. (1995) and Kustas and Norman (2000), discussed in Kustas et al. (2016), and shown below:

$$R_a = \frac{\ln\left(\frac{z_T - d_0}{z_{0H}}\right) - \Psi_h\left(\frac{z_T - d_0}{L}\right) + \Psi_h\left(\frac{z_{0H}}{L}\right)}{\kappa' u_*}$$
$$R_s = \frac{1}{c(T_S - T_A)^{1/3} + b u_s} \quad \text{(A4),}$$
$$R_x = \frac{C'}{LAI}\left(\frac{l_w}{U_{d_0 + z_{0M}}}\right)^{1/2}$$

where $u_*$ is the friction velocity (m $s^{-1}$) computed as:

$$u_* = \frac{\kappa' u}{[\ln(\frac{z_u - d_0}{z_{0M}}) - \Psi_m(\frac{z_u - d_0}{L}) + \Psi_m(\frac{z_{0M}}{L})]} \quad \text{(A5).}$$

20

Figure 4: Unmasked Page

$$\begin{aligned} R_n &\approx H + LE + G \\ R_{n,S} &\approx H_S + LE_S + G \\ R_{n,C} &\approx H_C + LE_C \end{aligned} \quad \text{(A1)}$$

$$\begin{aligned} H = H_C + H_S = & \quad \rho_{air} C_p \frac{T_{AC} - T_A}{R_a} \\ = & \quad \rho_{air} C_p [\frac{T_C - T_{AC}}{R_x} + \frac{T_S - T_{AC}}{R_s}] \end{aligned} \quad \text{(A2)}$$

$$T_{AC} = \frac{\frac{T_A}{R_a} + \frac{T_C}{R_x} + \frac{T_S}{R_S}}{\frac{1}{R_a} + \frac{1}{R_x} + \frac{1}{R_S}} \quad \text{(A3).}$$

$$\begin{aligned} R_a &= \frac{\ln\left(\frac{z_T - d_0}{z_{0H}}\right) - \Psi_h\left(\frac{z_T - d_0}{L}\right) + \Psi_h\left(\frac{z_{0H}}{L}\right)}{\kappa' \, u_*} \\ R_s &= \frac{1}{c(T_S - T_A)^{1/3} + b u_s} \\ R_x &= \frac{C'}{LAI} \left( \frac{l_w}{U_{d_0 + z_{0M}}} \right)^{1/2} \end{aligned} \quad \text{(A4),}$$

$$u_* = \frac{\kappa' u}{[\ln(\frac{z_u - d_0}{z_{0M}}) - \Psi_m(\frac{z_u - d_0}{L}) + \Psi_m(\frac{z_{0M}}{L})]} \quad \text{(A5).}$$

Figure 5: Masked Page

When the downstream prepare_element method in code_formula_model.py is called, it now uses this get_masked_image function instead of the standard get_image. By cropping the formula's bounding box from this pre-masked page, the system ensures that the final image snippet sent to the SmolDocling model contains only the formula itself, improving the signal-to-noise ratio and preventing recognition errors caused by textual artifacts.

#### 3.6.6 Heuristic-Based Merging of Fragmented Layout Clusters

A second challenge was identified in the output of the base layout analysis model: the fragmentation of single document elements into multiple, distinct bounding boxes. This issue was common with complex, multi-line mathematical formulas, which were often incorrectly segmented into several independent FORMULA clusters, as exemplified in Figure 6.

This fragmentation would lead to the downstream formula recognition model receiving only partial equations, resulting in incomplete and unusable LaTeX output.

To resolve this, a post-processing step was added within docling/utils/layout_postprocessor.py. The purpose of this step is to apply rule-based heuristics to refine and correct the raw output of the initial layout model before the final document structure is assembled.

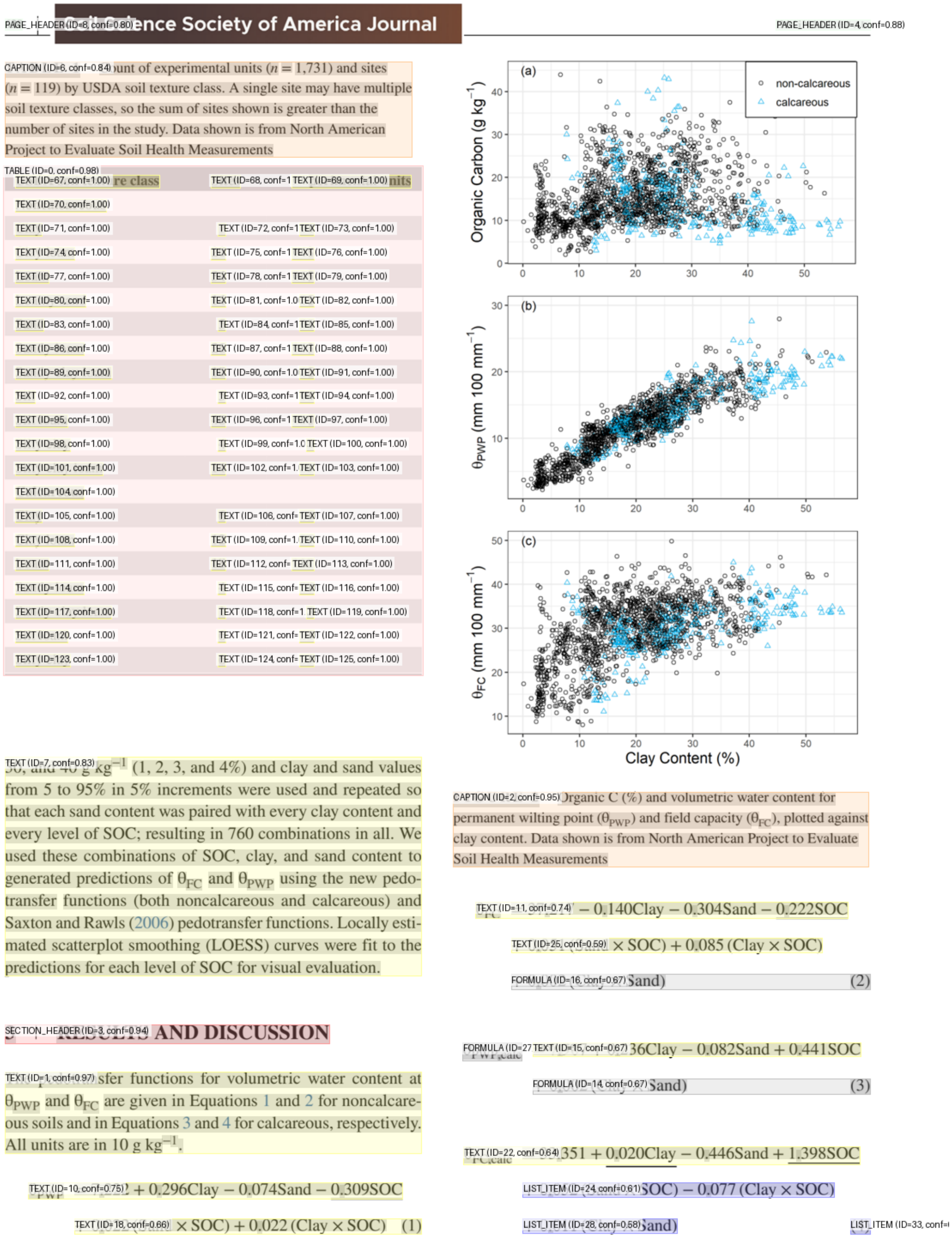

Figure 6: Fragmented Formulas

```python
def _merge_vertically_adjacent_formulas(
    self,
    clusters,
    vertical_threshold_factor=1.8,
    horizontal_overlap_threshold=0.7,
    padding=50,
    alignment_threshold=20,  # pixel-based threshold
    max_alignment_ratio=0.2,  # ratio-based threshold
):
    """
    Merge vertically adjacent FORMULA clusters using Union-Find.

    New logic:
    1) Compute dynamic vertical threshold based on median formula height
    2) Expand bounding boxes horizontally using padding to compute overlap
    3) Compute both absolute (pixel) differences and ratio-based alignment
    4) Extract formula numbers from cell texts. If both have numbers
       and differ, skip merge
    5) If neither has formula number, require very small vertical gap
       (<= 10 px) and nearly complete horizontal overlap (>= 0.9)
    6) Otherwise, use normal criteria based on vertical gap
    """
    def compute_alignment_factor(c1, c2):
        # Compute average width and return max normalized edge difference
        w1 = c1.bbox.r - c1.bbox.l
        w2 = c2.bbox.r - c2.bbox.l
        avg_width = max((w1 + w2) / 2.0, 1e-6)
        left_diff = abs(c1.bbox.l - c2.bbox.l)
        right_diff = abs(c1.bbox.r - c2.bbox.r)
        return max(left_diff / avg_width, right_diff / avg_width)

    # Only process clusters labelled as FORMULA
    formula_clusters = [c for c in clusters
                        if c.label == DocItemLabel.FORMULA]
    non_formula_clusters = [c for c in clusters
                            if c.label != DocItemLabel.FORMULA]
    if not formula_clusters:
        return clusters

    # Sort by the top coordinate
    formula_clusters.sort(key=lambda c: c.bbox.t)
    heights = [c.bbox.b - c.bbox.t for c in formula_clusters]
    median_height = np.median(heights) if heights else 0
    vertical_threshold = median_height * vertical_threshold_factor

    # Build mapping from cluster id to cluster instance
    id_to_cluster = {c.id: c for c in formula_clusters}
    uf = UnionFind(list(id_to_cluster.keys()))
    n = len(formula_clusters)

    for i in range(n):
        for j in range(i + 1, n):
            c1 = formula_clusters[i]
            c2 = formula_clusters[j]

            # Compute vertical gap (c2 is assumed to be below c1)
            vertical_gap = c2.bbox.t - c1.bbox.b
            if vertical_gap < 0 or vertical_gap > vertical_threshold:
```

```
                continue

            # Expand bounding boxes horizontally using given padding
            expanded_bbox_c1 = type(c1.bbox)(
                l=c1.bbox.l - padding,
                t=c1.bbox.t,
                r=c1.bbox.r + padding,
                b=c1.bbox.b
            )
            expanded_bbox_c2 = type(c2.bbox)(
                l=c2.bbox.l - padding,
                t=c2.bbox.t,
                r=c2.bbox.r + padding,
                b=c2.bbox.b
            )

            # Calculate horizontal overlap
            horizontal_overlap = (min(expanded_bbox_c1.r, expanded_bbox_c2.r) -
                                  max(expanded_bbox_c1.l, expanded_bbox_c2.l))
            min_width = min(expanded_bbox_c1.r - expanded_bbox_c1.l,
                               expanded_bbox_c2.r - expanded_bbox_c2.l)
            overlap_ratio = horizontal_overlap / min_width if min_width > 0 else 0

            left_diff = abs(c1.bbox.l - c2.bbox.l)
            right_diff = abs(c1.bbox.r - c2.bbox.r)
            alignment_factor = compute_alignment_factor(c1, c2)

            # Extract formula numbers using _extract_formula_number
            num1 = self._extract_formula_number(c1)
            num2 = self._extract_formula_number(c2)

            # Branch 4) If both have numbers and differ, skip
            if num1 and num2 and num1 != num2:
                continue

            # Normal merging criteria based on vertical gap and alignment
            # Pre-calculate default geometry-based threshold
            if vertical_gap <= (0.5 * vertical_threshold):
                if (left_diff <= alignment_threshold and
                    right_diff <= alignment_threshold):
                    required_overlap = horizontal_overlap_threshold
                else:
                    required_overlap = 0.85
            elif vertical_gap <= vertical_threshold:
                if alignment_factor <= max_alignment_ratio:
                    required_overlap = 0.85
                else:
                    continue
            else:
                continue

            # Branch handling based on formula numbers
            # Branch 2) Both None => require small gap and high overlap
            if num1 is None and num2 is None:
                if vertical_gap > 3:  # Small gap required
                    continue
                required_overlap = max(required_overlap, 0.9)
            # Branch 3) One number missing => stricter requirements
```

```
            elif (num1 is None) != (num2 is None):
                if vertical_gap > 12.8:
                    continue
                required_overlap = max(required_overlap, 0.95)

            # Final merge check
            if overlap_ratio >= required_overlap:
                uf.union(c1.id, c2.id)

    # Create merged clusters
    groups = uf.get_groups()
    merged_formula_clusters = []
    for group_ids in groups.values():
        group_clusters = [id_to_cluster[g] for g in group_ids]
        merged_bbox = type(group_clusters[0].bbox)(
            l=min(c.bbox.l for c in group_clusters),
            t=min(c.bbox.t for c in group_clusters),
            r=max(c.bbox.r for c in group_clusters),
            b=max(c.bbox.b for c in group_clusters)
        )
        combined_cells = []
        for c in group_clusters:
            combined_cells.extend(c.cells)

        merged_cluster = group_clusters[0]
        merged_cluster.bbox = merged_bbox
        merged_cluster.cells = self._sort_cells(
            self._deduplicate_cells(combined_cells)
        )
        merged_formula_clusters.append(merged_cluster)

    return non_formula_clusters + merged_formula_clusters
```

The core of this enhancement is the new _merge_vertically_adjacent_formulas method. This algorithm identifies and merges clusters that likely belong to the same formula. It operates on heuristics designed to distinguish between separate, adjacent equations and multiple lines of a single equation:

1. **Dynamic Proximity Threshold:** Rather than using a fixed pixel distance, the method first calculates the median height of all detected formulas on a page. It then defines a dynamic vertical proximity threshold based on a factor of this median height, allowing it to adapt to documents with different font sizes and line spacing.
2. **Formula Number Extraction:** A helper function, _extract_formula_number, uses regular expressions to find equation numbers (e.g., (1), (2a), (A5)) within the text of each formula cluster.
3. **Contextual Merging Logic:** The algorithm iterates through pairs of vertically adjacent formula clusters and applies a decision-making process:
    - If both clusters have distinct formula numbers (e.g., (1) and (2)), they are identified as separate equations and are not merged.
    - If neither cluster has a formula number, they are only merged if the vertical gap between them is minimal and their horizontal overlap is nearly complete, which is characteristic of multi-line equations without a single encompassing number.
    - If one cluster has a number and the adjacent one does not, a stricter set of proximity and overlap rules is applied.

Once candidate clusters for merging are identified, a Union-Find data structure is used to group them. The final step involves creating a new, single bounding box that encompasses all the merged clusters and combining their constituent text cells. This new, unified cluster is then passed to the next stage of the pipeline, ensuring the entire formula is processed as a single unit as demonstrated in Figure 7.

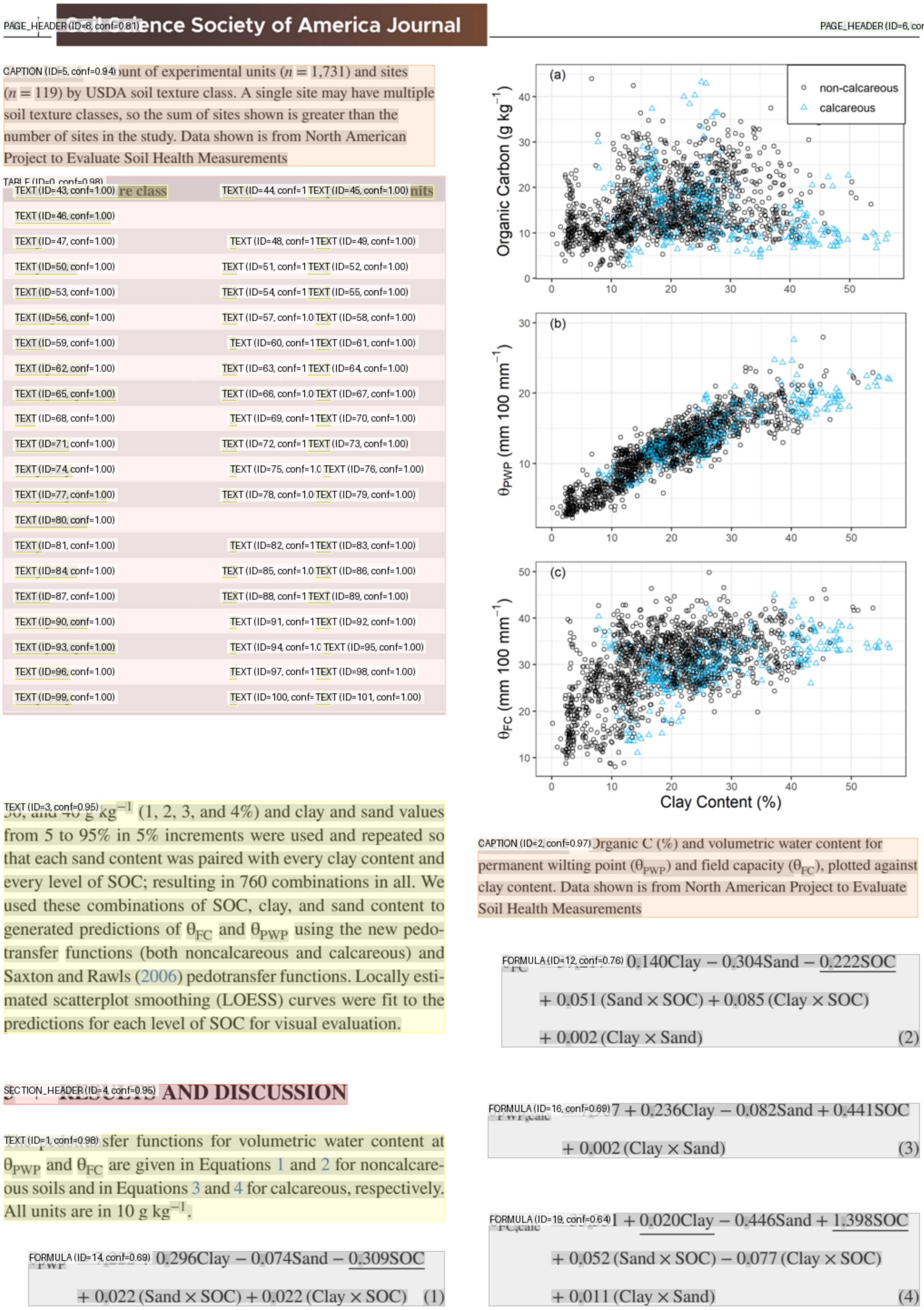

Figure 7: Unfragmented Formulas

Additional heuristics, such as the _filter_tables_containing_page_footer method, were also added to this module to correct other common layout analysis errors, improving the overall quality of the parsed document structure.

#### 3.6.7 Misclassification of Page Layouts

This section details further enhancements made to the layout post-processing logic to address recurring classification errors observed in certain document types, such as pre-prints or articles with line numbers.

A common failure mode was observed where the layout detection model would misclassify an entire page or large sections of a page as a single, large TABLE cluster. This issue was common in documents that featured line numbers running down the left-hand margin, as seen in Figure 8.

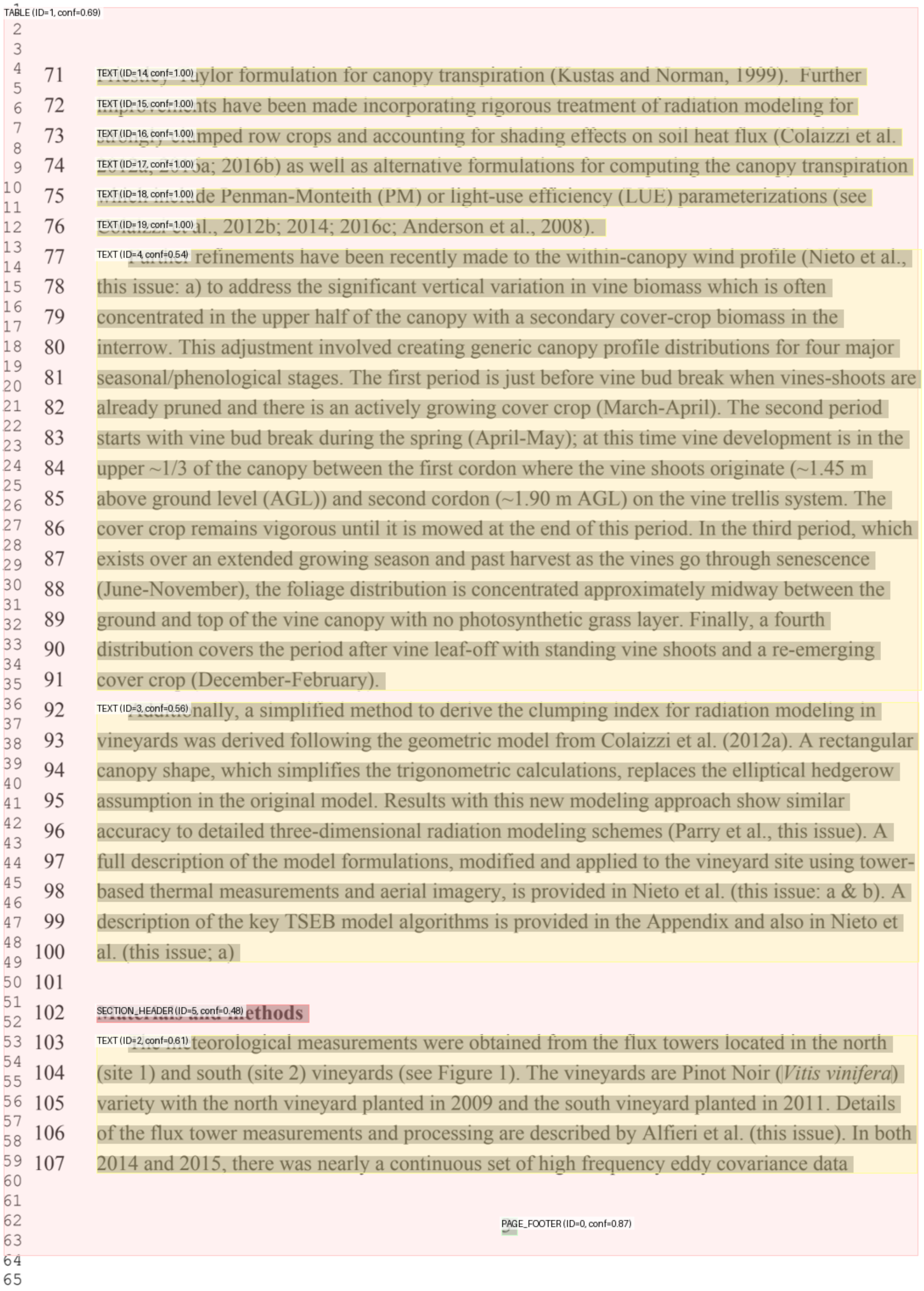

ylor formulation for canopy transpiration (Kustas and Norman, 1999). Further ts have been made incorporating rigorous treatment of radiation modeling for mped row crops and accounting for shading effects on soil heat flux (Colaizzi et al. a; 2016b) as well as alternative formulations for computing the canopy transpiration de Penman-Monteith (PM) or light-use efficiency (LUE) parameterizations (see al., 2012b; 2014; 2016c; Anderson et al., 2008).

refinements have been recently made to the within-canopy wind profile (Nieto et al., this issue: a) to address the significant vertical variation in vine biomass which is often concentrated in the upper half of the canopy with a secondary cover-crop biomass in the interrow. This adjustment involved creating generic canopy profile distributions for four major seasonal/phenological stages. The first period is just before vine bud break when vines-shoots are already pruned and there is an actively growing cover crop (March-April). The second period starts with vine bud break during the spring (April-May); at this time vine development is in the upper ~1/3 of the canopy between the first cordon where the vine shoots originate (~1.45 m above ground level (AGL)) and second cordon (~1.90 m AGL) on the vine trellis system. The cover crop remains vigorous until it is mowed at the end of this period. In the third period, which exists over an extended growing season and past harvest as the vines go through senescence (June-November), the foliage distribution is concentrated approximately midway between the ground and top of the vine canopy with no photosynthetic grass layer. Finally, a fourth distribution covers the period after vine leaf-off with standing vine shoots and a re-emerging cover crop (December-February).

nally, a simplified method to derive the clumping index for radiation modeling in vineyards was derived following the geometric model from Colaizzi et al. (2012a). A rectangular canopy shape, which simplifies the trigonometric calculations, replaces the elliptical hedgerow assumption in the original model. Results with this new modeling approach show similar accuracy to detailed three-dimensional radiation modeling schemes (Parry et al., this issue). A full description of the model formulations, modified and applied to the vineyard site using tower-based thermal measurements and aerial imagery, is provided in Nieto et al. (this issue: a & b). A description of the key TSEB model algorithms is provided in the Appendix and also in Nieto et al. (this issue; a)

**ethods**

teorological measurements were obtained from the flux towers located in the north (site 1) and south (site 2) vineyards (see Figure 1). The vineyards are Pinot Noir (*Vitis vinifera*) variety with the north vineyard planted in 2009 and the south vineyard planted in 2011. Details of the flux tower measurements and processing are described by Alfieri et al. (this issue). In both 2014 and 2015, there was nearly a continuous set of high frequency eddy covariance data

Figure 8: Table Missclassification

The presence of this repeating, structured numerical column appeared to mislead the model into perceiving the entire text block as tabular data. This resulted in the loss of all other semantic distinctions for that page (e.g., paragraphs, headers, lists), rendering the content unusable for downstream extraction.

#### 3.6.8 Heuristic-Based Re-classification of Misidentified Clusters

To correct this, a new heuristic-based filtering method, _filter_tables_containing_page_footer, was added to the main postprocess pipeline. While the method name suggests a focus on page footers, its logic was designed to identify and re-classify any cluster that was erroneously labeled as a table-like structure when it was, in fact, the main body text of the page.

```python
# In scripts/docling/utils/layout_postprocessor.py

#| label: filter-large-tables
#| eval: false

def _filter_tables_containing_page_footer(
    self,
    clusters: List[Cluster],
    min_area_ratio: float = 0.70,
    min_cells_threshold: int = 50,
    min_density_threshold: float = 0.001
) -> List[Cluster]:
    """
    Enhanced logic to avoid reclassifying legitimate large tables.
    """
    large_labels = {
        DocItemLabel.TABLE,
        DocItemLabel.DOCUMENT_INDEX,
        DocItemLabel.KEY_VALUE_REGION,
        DocItemLabel.FORM
    }
    page_area = self.page_size.width * self.page_size.height
    reclassified_count = 0

    for cluster in clusters:
        if cluster.label in large_labels:
            cluster_area_ratio = cluster.bbox.area() / page_area
            if cluster_area_ratio >= min_area_ratio:
                all_cells = self._collect_all_cells(cluster)
                cell_density = len(all_cells) / cluster.bbox.area()

                # Reclassify if cluster is large but sparsely populated
                if (len(all_cells) < min_cells_threshold or
                    cell_density < min_density_threshold):
                    cluster.label = DocItemLabel.TEXT
                    reclassified_count += 1

    return clusters
```

The algorithm operates on heuristics to distinguish between a legitimate, large table and a misclassified text block:

1. **Target Selection:** The function first identifies any cluster labeled as a TABLE (or other large "wrapper" types like FORM) for inspection.
2. **Area Threshold:** It only considers clusters that are large, occupying a significant portion of the page (e.g., min_area_ratio of 70%). This prevents the algorithm from affecting smaller, legitimate tables.
3. **Cell Count & Density Check:** This is the core heuristic. It recursively collects all the individual text cells within the large cluster and calculates two metrics: the absolute number of cells and the "cell density" (number of cells per unit of bounding box area).

4. **Re-classification Logic:** If a cluster is very large but contains a low number of text cells or has a very low cell density, it is unlikely to be a real table. In such cases, the algorithm re-classifies the cluster's label from TABLE to TEXT.

By adding this function to the postprocess pipeline, the system can now identify and relabel these large, sparse, misclassified text blocks, preserving the semantic integrity of the page structure as demonstrated by the corrected output in Figure 9.

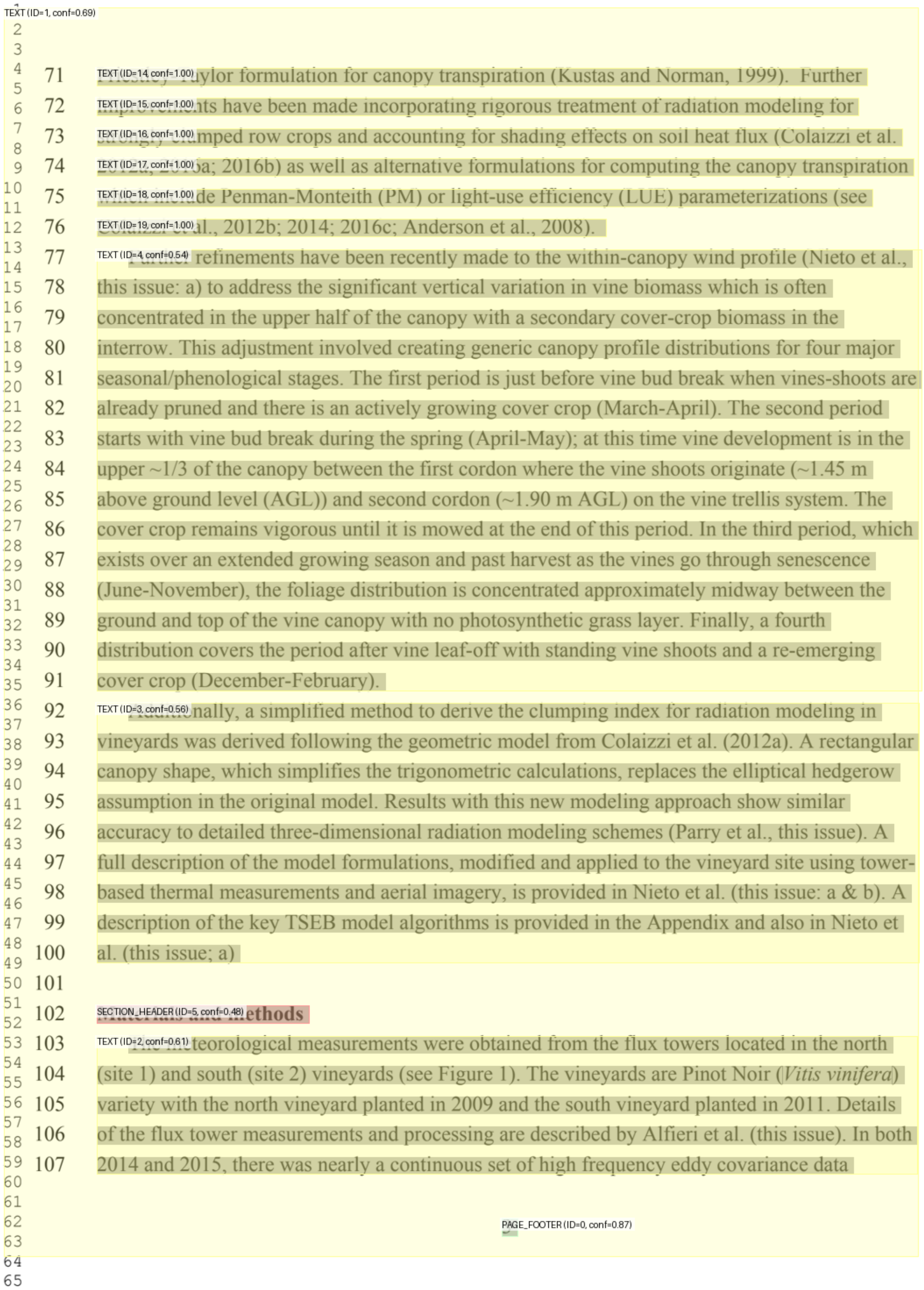

...ylor formulation for canopy transpiration (Kustas and Norman, 1999). Further ...ts have been made incorporating rigorous treatment of radiation modeling for ...mped row crops and accounting for shading effects on soil heat flux (Colaizzi et al. ...a; 2016b) as well as alternative formulations for computing the canopy transpiration ...de Penman-Monteith (PM) or light-use efficiency (LUE) parameterizations (see ...l., 2012b; 2014; 2016c; Anderson et al., 2008).

... refinements have been recently made to the within-canopy wind profile (Nieto et al., this issue: a) to address the significant vertical variation in vine biomass which is often concentrated in the upper half of the canopy with a secondary cover-crop biomass in the interrow. This adjustment involved creating generic canopy profile distributions for four major seasonal/phenological stages. The first period is just before vine bud break when vines-shoots are already pruned and there is an actively growing cover crop (March-April). The second period starts with vine bud break during the spring (April-May); at this time vine development is in the upper ~1/3 of the canopy between the first cordon where the vine shoots originate (~1.45 m above ground level (AGL)) and second cordon (~1.90 m AGL) on the vine trellis system. The cover crop remains vigorous until it is mowed at the end of this period. In the third period, which exists over an extended growing season and past harvest as the vines go through senescence (June-November), the foliage distribution is concentrated approximately midway between the ground and top of the vine canopy with no photosynthetic grass layer. Finally, a fourth distribution covers the period after vine leaf-off with standing vine shoots and a re-emerging cover crop (December-February).

...nally, a simplified method to derive the clumping index for radiation modeling in vineyards was derived following the geometric model from Colaizzi et al. (2012a). A rectangular canopy shape, which simplifies the trigonometric calculations, replaces the elliptical hedgerow assumption in the original model. Results with this new modeling approach show similar accuracy to detailed three-dimensional radiation modeling schemes (Parry et al., this issue). A full description of the model formulations, modified and applied to the vineyard site using tower-based thermal measurements and aerial imagery, is provided in Nieto et al. (this issue: a & b). A description of the key TSEB model algorithms is provided in the Appendix and also in Nieto et al. (this issue; a)

...ethods

...teorological measurements were obtained from the flux towers located in the north (site 1) and south (site 2) vineyards (see Figure 1). The vineyards are Pinot Noir (*Vitis vinifera*) variety with the north vineyard planted in 2009 and the south vineyard planted in 2011. Details of the flux tower measurements and processing are described by Alfieri et al. (this issue). In both 2014 and 2015, there was nearly a continuous set of high frequency eddy covariance data

Figure 9: Table Correction

#### 3.6.9 Pre-emptive Filtering of Page-Level Artifacts

Further enhancements were added upstream in the docling/backend/docling_parse_v4_backend.py module to improve the initial quality of the data fed to the layout analysis model. This module serves as a low-level backend that interacts directly with the docling-parse C++ library, which is responsible for the initial extraction of raw text cells and their coordinates from the PDF document. It was observed that certain document formats, particularly pre-prints, often contain page-level artifacts like line numbers in the margins. These structured, non-substantive elements were found to frequently mislead the layout model, leading to classification errors where an entire page of text would be misidentified as a single, large TABLE cluster.

To mitigate this failure mode, the primary modification was made to the get_text_cells method within the DoclingParseV4PageBackend class. The original implementation of this method simply transformed the coordinate system of all extracted text cells and returned the complete, unfiltered list. The enhanced version introduces a new helper method, _is_left_margin_line_number, which applies geometric heuristics to identify and pre-emptively filter these line-number artifacts before they are passed to the layout model.

The heuristic function determines if a given text cell is a line number by evaluating three spatial properties. First, it confirms the cell is located within a narrow vertical band on the far-left of the page, defined by a LEFT_MARGIN_THRESHOLD (e.g., the leftmost 8% of the page width). Second, it verifies that the cell's bounding box width is less than this MAX_WIDTH_THRESHOLD, characteristic of short numerical strings. Finally, it ensures the cell has a reasonable height via MIN_HEIGHT_THRESHOLD to avoid incorrectly filtering other small page markings like footnote symbols. A cell is only flagged and removed if it satisfies all three conditions. By filtering the cell list with this function, the layout model receives a cleaner representation of the page's semantic content, reducing classification errors and preserving the integrity of the document structure for all downstream processing.

**Original Implementation (get_text_cells)**

```
def get_text_cells(self) -> Iterable[TextCell]:
    page_size = self.get_size()

    # Applies coordinate transformation but returns all cells
    [tc.to_top_left_origin(page_size.height) for tc in self._dpage.textline_cells]

    return self._dpage.textline_cells
```

**Modified Implementation with Heuristic Filtering**

```
def get_text_cells(self) -> Iterable[TextCell]:
    page_size = self.get_size()
    for tc in self._dpage.textline_cells:
        tc.to_top_left_origin(page_size.height)

    # Filter out cells identified as line numbers by the heuristic
    filtered_cells = [
        cell for cell in self._dpage.textline_cells
        if not self._is_left_margin_line_number(cell, page_size)
    ]
    return filtered_cells

def _is_left_margin_line_number(self, cell: TextCell, page_size: Size) -> bool:
    """
    Identifies if a cell is a line number based on geometric properties.
    """
    LEFT_MARGIN_THRESHOLD = 0.08
    MIN_HEIGHT_THRESHOLD = 5
    MAX_WIDTH_THRESHOLD = page_size.width * LEFT_MARGIN_THRESHOLD

    bbox = cell.rect.to_bounding_box()

    is_within_left_margin = bbox.l < MAX_WIDTH_THRESHOLD
    is_small_horizontal = bbox.width < MAX_WIDTH_THRESHOLD
    is_not_too_short = bbox.height >= MIN_HEIGHT_THRESHOLD # Corrected logic

    return is_within_left_margin and is_small_horizontal and is_not_too_short
```

#### 3.6.10 Refinements to the Underlying Layout Predictor

Targeted modifications were also made to the underlying docling_ibm_models/layoutmodel/layout_predictor.py module. This module is responsible for the initial, low-level detection of layout elements on a page image by executing the RTDetrForObjectDetection model. While the primary logic of the model itself was not altered, refinements were made to its data handling and execution to ensure high-performance and stable operation within the custom parallelized framework.

The most critical modification addresses a performance bottleneck related to hardware utilization. In the predict method, the target_sizes tensor, which is required by the post_process_object_detection function to correctly rescale bounding boxes to the original image dimensions was originally created on the CPU by default.

```python
@torch.inference_mode()
def predict(self, orig_img: Union[Image.Image, np.ndarray]) -> Iterable[dict]:
    results = self._image_processor.post_process_object_detection(
        outputs,
        target_sizes=torch.tensor([page_img.size[::-1]]), # Tensor created on CPU by default
        threshold=self._threshold,
    )
    w, h = page_img.size
    result = results[0]
    for score, label_id, box in zip(result["scores"], result["labels"], result["boxes"]):
# ...
        # Manual clamping of coordinates
        l = min(w, max(0, bbox_float[0]))
        t = min(h, max(0, bbox_float[1]))
        r = min(w, max(0, bbox_float[2]))
        b = min(h, max(0, bbox_float[3]))
        yield { # ...
        }
```

When the model was running on a GPU, this mismatch forced an expensive and unnecessary cross-device data transfer during post-processing. The implementation was corrected to explicitly create this tensor on the same device as the model (device=self._device), eliminating the synchronization penalty and ensuring more efficient CUDA utilization.

```python
# ...
@torch.inference_mode()
def predict(self, orig_img: Union[Image.Image, np.ndarray]) -> Iterable[dict]:
# ...
    # Explicitly create the tensor on the same device as the model (e.g., CUDA)
    target_sizes = torch.tensor([[original_height, original_width]], device=self._device)
    results = self._image_processor.post_process_object_detection(
        outputs, target_sizes=target_sizes, threshold=self._threshold
    )
    result = results[0]
    # Directly use the results after converting to NumPy arrays
    boxes = result["boxes"].cpu().numpy()
    scores = result["scores"].cpu().numpy()
    labels = result["labels"].cpu().numpy()
    for score, label_id, box in zip(scores, labels, boxes):
# ...
        # No more manual clamping
        l, t, r, b = box
        yield { # ...
        }
```

Additionally, the code was modernized by removing redundant manual logic for clamping bounding box coordinates. The original implementation manually ensured that the box coordinates did not exceed the page dimensions after being returned by the post-processing function. The refined version removes this step, delegating the responsibility for coordinate clamping to the transformers library's post_process_object_detection function. This change makes the code cleaner, more maintainable, and adheres more closely to the intended use of the library's API. While subtle, these enhancements to the core predictor were essential for ensuring the stability and high throughput required by the system's parallel architecture.

#### 3.6.11 Final Implementation in the ETL Pipeline

The re-engineered Docling system is integrated as a distinct stage in the ETL pipeline, handled by the do_docling_extraction function.

```python
# In scripts/docling_multi_mp_gui.py

#| label: docling-extraction-orchestrator
#| eval: false

def do_docling_extraction(
    df: pd.DataFrame,
    progress_callback=None
) -> pd.DataFrame:
    """
    Processes DataFrame rows in parallel using config settings.
    """
    global MAX_WORKERS_DOCLING
    num_records = len(df)
    # Use the worker count from config, ensuring it's at least 1
    max_workers_to_use = max(1, MAX_WORKERS_DOCLING)

    output_cols = ["FullText", "TablesJson", "EquationsJson",
                   "TokenCount", "Error"]
    for col in output_cols:
        if col not in df.columns:
            if col == "TokenCount":
                df[col] = 0
            elif col == "Error":
                df[col] = None
            else:
                df[col] = ""  # Default to empty string for text/JSON columns

    futures = {}
    processed_count = 0

    with ProcessPoolExecutor(
        max_workers=max_workers_to_use,
        initializer=worker_initializer
    ) as executor:
        for idx, row in df.iterrows():
            pdf_path = row.get("PDFPath", "")
            if not pdf_path or not os.path.exists(pdf_path):
                df.loc[idx, output_cols] = [
                    "PDF_PATH_ERROR", "[]", "[]", 0, "PDF_PATH_ERROR"
                ]
                processed_count += 1
                continue

            future = executor.submit(extract_pdf_with_docling, pdf_path)
            futures[future] = idx  # Map future to DataFrame index

        # Process results as they complete
        for future in as_completed(futures):
            irow = futures[future]  # Get the original DataFrame index
            try:
                result = future.result()  # Get the dict returned by worker
                # Update DataFrame using .loc with index 'irow'
                for col in output_cols:
                    df.loc[irow, col] = result.get(col)
```

```
            except Exception as e:
                df.loc[irow, output_cols] = [
                    "FUTURE_ERROR", "[]", "[]", 0,
                    f"FutureError: {type(e).__name__}"
                ]

    return df
```

This function receives a pandas DataFrame containing paths to the downloaded PDF articles. It distributes the processing of each PDF to the pool of GPU-powered workers. Each worker executes the extract_pdf_with_docling function, which initializes a dedicated DocumentConverter instance configured for its assigned GPU. The worker processes its assigned PDF, extracts the full text (exported as Markdown to preserve structure), and identifies and serializes all tabular data and mathematical formulas into JSON arrays. Upon completion, the structured outputs, FullText, TablesJson, EquationsJson, along with a TokenCount are returned to the main process and integrated back into the corresponding row of the DataFrame. The resulting DataFrame is then saved as a new artifact (output.feather), ready for the subsequent LLM-based field extraction stage.

### 3.7 Stage 5: Structured Field Extraction using Large Language Models

Following the extraction of raw text via document layout analysis, the pipeline proceeds to the structured field extraction stage, handled by the scripts/field_extraction.py module. The objective of this component is to parse the unstructured text from each article and populate a predefined set of structured data fields. This process transforms dense, narrative content into a queryable, machine-readable format suitable for populating the knowledge graph and facilitating analysis.

#### 3.7.1 Methodology: Iterative, Context-Aware Extraction with Local LLMs

The system uses an extraction strategy centered around a locally deployed Large Language Model (Qwen/Qwen3-32B). This self-hosted approach, leveraging the project's multi-GPU hardware, ensures data privacy, eliminates reliance on external API costs and latency, and provides control over the inference process. The extraction is not a single pass; rather, it is an iterative process designed to build a cumulative record for each document.

The core of the methodology is the _process_one_row function, which operates on each article's full text. To manage the extensive length of academic papers and stay within the LLM's context window, the chunk_text_for_extraction utility first splits the full text into overlapping chunks of a configured token size (e.g., 8000 tokens with a 500-token overlap). The system then iterates through these chunks, performing an LLM call for each one.

A key innovation is the use of a dynamic, context-aware prompt. For each chunk, the LLM is provided not only with the text of that chunk but also with the current state of the metadata that has been extracted from all previous chunks of the same document. This allows the model to incrementally enrich the data, fill in missing fields, and use previously extracted information as context for interpreting the current text chunk. The prompt dynamically generates a list of fields to be extracted based on the SELECTED_COLS defined in the etl_config.json file, which are interactively selected via the GUI. Rather than just providing the field name (e.g., "LCA System Boundaries"), the prompt includes a detailed explanation from the FIELD_EXPLANATIONS dictionary, guiding the LLM with a precise definition of the information to look for (e.g., "Describe the scope and boundaries of any Life Cycle Assessment (LCA) mentioned, such as cradle-to-grave, cradle-to-gate…").

```
# In scripts/field_extraction.py, illustrating dynamic prompt generation

# ... inside _process_one_row ...

# The current state of extracted data is passed as context
metadata_context_json = json.dumps(final_merged_data, indent=2)

# The prompt is formatted with the context, the new text chunk,
# and the list of fields with their detailed explanations.
prompt = extraction_template.format(
    metadata_json=metadata_context_json,
    chunk_text=chunk_txt,
    field_list_placeholder=field_list_str
)

# ... LLM call is made with this enhanced prompt ...
```

#### 3.7.2 Parallelized Execution and Data Merging

To handle the processing of hundreds of articles, the entire operation is parallelized. The extract_additional_fields function uses a ThreadPoolExecutor to process multiple articles (rows in the DataFrame) concurrently. Within each of these threads, a nested ThreadPoolExecutor is used to process the individual text chunks for that article in parallel. This multi-level parallelism ensures utilization of the available hardware.

The output from each LLM call is a JSON object containing only the fields found within that specific text chunk. A merging function, unify_fields, is then used to integrate this new data into the master record for the article. This function handles various data types, appending items to list-based fields (e.g., Pollutant Terms), overwriting simple text fields with more specific information, and aggregating complex structured data like the Metrics field. To prevent performance degradation from repeated LLM calls on identical text, a cache (llm_response_cache) stores the output for each unique prompt, returning the cached result if the same text chunk is ever processed again. This entire process, from chunking to parallelized inference and merging, results in a detailed and structured dataset, which is then saved as an enriched Feather file, ready for the final stages of the ETL pipeline.

### 3.8 Stage 6: Thematic Analysis via Topic Modeling

Following the extraction of structured data, the pipeline performs a thematic analysis of the textual corpus to identify latent topics, which serves as a method for research gap analysis. This is achieved through the implementation of a configurable and parallelized Latent Dirichlet Allocation (LDA) topic modeling workflow, handled by the scripts/topic_modeling_gui.py module. The objective is to distill the unstructured text from hundreds of articles into a set of coherent, interpretable topics whose prevalence and relationships can be quantitatively assessed.

#### 3.8.1 Methodology: Probabilistic Topic Modeling with LDA

The core of this stage is Latent Dirichlet Allocation (LDA), a generative probabilistic model for discrete data. The assumption of LDA is that each document in a corpus is a mixture of various topics, and each topic is a distribution of words. The model does not know what the topics are in advance; it learns them by analyzing the patterns of word co-occurrence across the entire set of documents.

Mathematically, LDA models a document as being generated by the following process:

1. For each document $(d)$ in the corpus $(D)$, choose a distribution over topics $(\theta_d \sim \text{Dir}(\alpha))$.
2. For each word $(w_n)$ in document $(d)$:
    a. Choose a topic $(z_n \sim \text{Categorical}(\theta_d))$.
    b. Choose a word $(w_n)$ from $(p(w_n \mid z_n, \beta))$, the probability of word $(w_n)$ given topic $(z_n)$.

The goal of the model training is to infer the hidden variables: the topic distributions per document $(\theta_d)$ and the word distributions per topic $(\beta_k)$. The system uses scikit-learn's implementation of LDA, which employs a variational Bayes algorithm to approximate these posterior distributions. The output for each topic $(k)$ is a probabilistic representation described as a list of words that are most likely to belong to that topic.

#### 3.8.2 Data Preprocessing and Hyperparameter Optimization

The quality of an LDA model depends on both the cleanliness of the input text and the choice of model hyperparameters. The pipeline therefore begins with a parallelized text preprocessing workflow (preprocess_pipeline function) that tokenizes, removes stopwords, and lemmatizes the text from each document.

To select the optimal hyperparameters for the corpus, the system implements a parallelized grid search, handled by the _search_best_lda_params function. The user can define the search space in the etl_config.json file interactively via the GUI, specifying ranges for key parameters:

- **num_topics (k)**: The number of latent topics to discover. The system searches over a dynamic range to find the optimal granularity.
- **passes / iterations**: The number of passes the algorithm makes over the corpus during training.
- **n-gram Parameters (bigram_threshold, trigram_threshold)**: These control the formation of common multi-word phrases (e.g., "cover crop," "random forest model"), treating them as single tokens to generate more coherent topics.
- **Dictionary Filtering (no_below, no_above)**: These parameters prune the vocabulary by removing terms that are either too rare (no_below) or too common (no_above) to be thematically useful.

The grid search trains multiple LDA models in parallel using ProcessPoolExecutor, and each resulting model is evaluated using a Topic Coherence score.

#### 3.8.3 Model Evaluation: Topic Coherence

Topic coherence measures the degree of semantic similarity between the high-scoring words within a topic, providing a quantitative way to assess how interpretable a topic is. This system uses the $(C_v)$ coherence measure, which is based on a sliding window and the normalized pointwise mutual information (NPMI) of word pairs.

For a set of top $(N)$ words $(w_1, w_2, \ldots, w_N)$ in a given topic, the $(C_v)$ score is calculated as the average NPMI of all unique word pairs:

$$\text{Coherence}(V) = \frac{1}{\binom{N}{2}} \sum_{i=1}^{N-1} \sum_{j=i+1}^{N} \text{NPMI}(w_i, w_j)$$

where the Normalized Pointwise Mutual Information (NPMI) is given by:

$$\text{NPMI}(w_i, w_j) = \frac{\log \frac{P(w_i, w_j)}{P(w_i)P(w_j)}}{-\log P(w_i, w_j)}$$

Here, $P(w)$ is the probability of seeing word $w$ in a document, and $P(w_i, w_j)$ is the probability of seeing both words in the same document. A score closer to 1 indicates a more coherent topic. The grid search selects the hyperparameter set that produces the model with the highest average $C_v$ score across all its topics. Additionally, the calculate_topic_coherences function is used to compute the coherence score for each individual topic from the best model, allowing for assessment of the quality of each thematic cluster.

#### 3.8.4 Interpretation and Visualization

After the best model is identified, the system assigns a dominant topic to each document and uses an LLM (Qwen/Qwen3-32B) to generate a concise, human-readable label for each topic based on its top keywords. Finally, the generate_lda_visualization.py script uses the pyLDAvis library to create an interactive visualization. This plot maps the topics into a 2D space, where the size of each topic's circle represents its prevalence in the corpus and the distance between circles indicates their semantic dissimilarity. This provides a visual tool for identifying prevalent themes and discovering potentially under-represented research areas.

### 3.9 Stage 7: The Semantic Unification Pipeline

This section details the methodology used to resolve terminological ambiguity within the data extracted by the ETL pipeline, ensuring ontological consistency before ingestion into the knowledge graph. This core functionality is encapsulated within the scripts/dictionaries_gui.py module.

#### 3.9.1 The Challenge of Terminological Heterogeneity in Automated Extraction

A primary challenge in the automated processing of scientific literature arises from terminological heterogeneity. The field_extraction.py module, which uses a Large Language Model (LLM), extracts unstructured textual strings that often exhibit variation despite referring to the same underlying concept. For example, within the "Tillage Practices" field, semantically equivalent concepts may be described as "no-till," "zero tillage," or "direct drilling." To construct a coherent and queryable knowledge graph, as handled by kg_pipeline_gui.py, it is necessary that these synonymous variations are resolved into a single, canonical entity. The system addresses this challenge not through string matching or rule-based heuristics, but through a process of semantic unification, which quantifies the contextual meaning of extracted terms. This core functionality is encapsulated within the scripts/dictionaries_gui.py module.

#### 3.9.2 A Vector Space Model for Semantic Representation

The theory underpinning the unification process is the representation of language within a high-dimensional vector space, often referred to as word or phrase embedding. This approach posits that the meaning of a term can be captured by a dense numerical vector, where terms with similar meanings are located closer to each other in this geometric space. This system employs a pre-trained Sentence Transformer model (sentence-transformers/all-MiniLM-L6-v2) to perform this transformation. This model generates a 384-dimensional vector for any given textual input, mapping each term to a unique coordinate in a 384-dimensional semantic space.

#### 3.9.3 Pre-computation of the Canonical Knowledge Base

The unification process is implemented through a sequence of operations designed for both accuracy and computational efficiency. At pipeline initialization, the precompute_all function within dictionaries_gui.py is executed.

```
def precompute_all(
    config_path="dictionaries_config.json",
    device: str = DEVICE_STR
):
    """
    Loads synonym dicts from config, then builds embeddings for each.
    """
    global _precomputed, _synonym_dictionaries, _precomputed_embeddings
    if _precomputed:
        return

    # Load synonym dicts first
    _synonym_dictionaries = load_synonym_dictionaries(config_path)
    if _synonym_dictionaries is None:
         raise ValueError("Synonym dictionaries failed to load.")

    model = get_st_model(device=device)
    # Initialize the dictionary to store computed embeddings
    _precomputed_embeddings = {}

    logger.info(f"Starting embedding precomputation for "
                f"{len(_synonym_dictionaries)} dictionaries...")

    # Dynamically build embeddings based on loaded dicts
    for dict_key, syn_dict in _synonym_dictionaries.items():
        logger.info(f"Processing dictionary: {dict_key}")
        if syn_dict:  # Check if dictionary is not empty
            phrases, embeds, label_map = build_candidate_embeddings(
                syn_dict, model=model, device=device
            )
            if embeds is not None:  # Check if generation was successful
                 _precomputed_embeddings[dict_key] = {
                      "phrases": phrases,
                      "embeds": embeds.to(device),  # Ensure correct device
                      "label_map": label_map
                 }
                 logger.info(f"Finished {dict_key}, {len(phrases)} phrases, "
                            f"embedding shape: {embeds.shape}")
            else:
                 logger.error(f"Embedding generation failed for "
                             f"dictionary: {dict_key}")
        else:
             logger.warning(f"Skipping empty synonym dictionary: {dict_key}")

    _precomputed = True
    logger.info(f"All configured synonym embeddings pre-computed and "
                f"stored on device '{device}'.")
```

This function iterates through the synonym dictionaries defined in config/dictionaries_config.json. For each dictionary, such as TILLAGE_PRACTICES_SYNONYMS or ML_AI_METHODS_SYNONYMS, it compiles a list of all canonical terms and their associated synonyms. Each of these phrases is then passed through the Sentence Transformer model to generate its corresponding 384-dimensional embedding. The resulting collection of vectors and a mapping that links each vector back to its canonical parent term are cached in memory. This pre-computation creates a static, numerically-indexed semantic map of the entire known vocabulary for each category, which allows subsequent matching operations to be performed as numerical comparisons, rather than repeated, computationally expensive model inferences.

#### 3.9.4 Quantifying Semantic Similarity: The Cosine Similarity Metric

Once the field_extraction.py module provides a new, unstructured term for a given field (e.g., "conservation tillage" for the "Tillage Practices" category), the term is first converted into a 384-dimensional query vector, $q$. The _find_best_match function then compares this query vector to every pre-computed candidate vector, $c$, within the relevant semantic map.

```
#| label: best-match
#| eval: false

def _find_best_match(query_emb, dict_key):
    """Internal helper to find best match using precomputed embeddings."""
    global _precomputed_embeddings
    if not _precomputed:
        logger.error(f"Embeddings not precomputed. Cannot unify for "
                     f"key '{dict_key}'. Call precompute_all() first.")
        return None, 0.0

    if dict_key not in _precomputed_embeddings:
        logger.warning(f"No precomputed embeddings found for "
                       f"key '{dict_key}'.")
        return None, 0.0

    data = _precomputed_embeddings[dict_key]
    embeds = data.get("embeds")
    label_map = data.get("label_map")
    phrases = data.get("phrases")  # For logger

    if embeds is None or label_map is None or phrases is None:
         logger.error(f"Precomputed data is incomplete for "
                      f"key '{dict_key}'.")
         return None, 0.0

    try:
        from sentence_transformers import util
        scores = util.cos_sim(query_emb, embeds)[0]
        best_score = float(scores.max())
        best_idx = int(scores.argmax())
        return best_idx, best_score
    except Exception as sim_err:
        logger.error(f"Error during cosine similarity calculation for "
                     f"key '{dict_key}': {sim_err}", exc_info=True)
        return None, 0.0
```

The metric used for this comparison is the Cosine Similarity, which measures the cosine of the angle ($\theta$) between the two vectors and serves as a measure of their orientation and semantic alignment, independent of their magnitude. The formula for this calculation is:

$$\text{sim}(\mathbf{q}, \mathbf{c}) = \cos\theta = \frac{\mathbf{q} \cdot \mathbf{c}}{\|\mathbf{q}\| \, \|\mathbf{c}\|} = \frac{\sum_{i=1}^{n} q_i \, c_i}{\sqrt{\sum_{i=1}^{n} q_i^2} \sqrt{\sum_{i=1}^{n} c_i^2}}$$

where:

- $n$ is the embedding dimension (384 for **all-MiniLM-L6-v2**);
- $\mathbf{q} \in \mathbb{R}^n$ is the 384-dimensional vector representing the new, unstructured query term;
- $\mathbf{c} \in \mathbb{R}^n$ is a 384-dimensional candidate vector from the pre-computed knowledge base;

- $q_i$ and $c_i$ are the $i$-th components of the query and candidate vectors, respectively.

The cosine similarity ranges from $-1$ (opposite) to $1$ (identical), where $1$ signifies that the vectors point in the exact same direction (a perfect semantic match), $0$ indicates they are orthogonal (semantically unrelated), and $-1$ indicates they are diametrically opposed.

#### 3.9.5 Threshold-Based Mapping for Ontological Consistency

After calculating the similarity score between the query vector and all candidate vectors, the system identifies the maximum score, representing the "closest" known term in the semantic space. The final step of the unification process is to transform this probabilistic similarity score into a deterministic mapping. This is achieved by comparing the highest score against a pre-defined confidence threshold (e.g., 0.55). If the score is greater than or equal to this threshold, the match is accepted as valid, and the system uses the pre-computed label_map to retrieve the canonical term associated with the best-matching vector. This canonical term is then used for ingestion into the knowledge graph. If the highest score falls below the threshold, the query term is considered a non-match, ensuring that ambiguous or out-of-domain terms are rejected, thereby safeguarding the ontological integrity and consistency of the final knowledge graph.

#### 3.9.6 Knowledge Graph Integration

The final stage of the ETL pipeline transforms the processed and enriched tabular data into a highly interconnected knowledge graph using a Neo4j graph database. The objective is to model the extracted metadata not as isolated rows, but as a network of distinct entities (such as articles, authors, diseases, and methods) and the explicit relationships that connect them. This graph structure enables complex, multi-faceted queries that would be inefficient or impossible with standard relational tables. The entire process, from data preparation to graph ingestion, is handled by scripts/kg_pipeline_gui.py and is driven by a declarative schema defined in the config/kg_pipeline.json file.

#### 3.9.7 Dynamic Graph Construction

The construction of the knowledge graph is a dynamic, two-step process for each article processed by the pipeline:

1. **Article Node Creation:** The process begins by creating or merging a central :Article node, using the article's Digital Object Identifier (DOI) as its unique key. All primary metadata, such as the title, citation count, Zotero key, and the results of the topic modeling stage, are set as properties on this main node.
2. **Entity and Relationship Mapping** The create_or_update_kg function then iterates through the field_mappings defined in the configuration file. For each field in the dataset (e.g., ml_methods_used), the system creates MERGE statements in the Cypher query language. For instance, for an article that used "Random Forest", the system ensures a node (:MLMethod {name: 'Random Forest'}) exists and then creates a [:USES_ML_METHOD] relationship from the :Article node to it. This procedure is applied across all mapped fields, creating a network of interconnected entities such as :HeartDisease, :PollutantTerm, and :StudyType, as visualized in the graph schema (Figure 10).

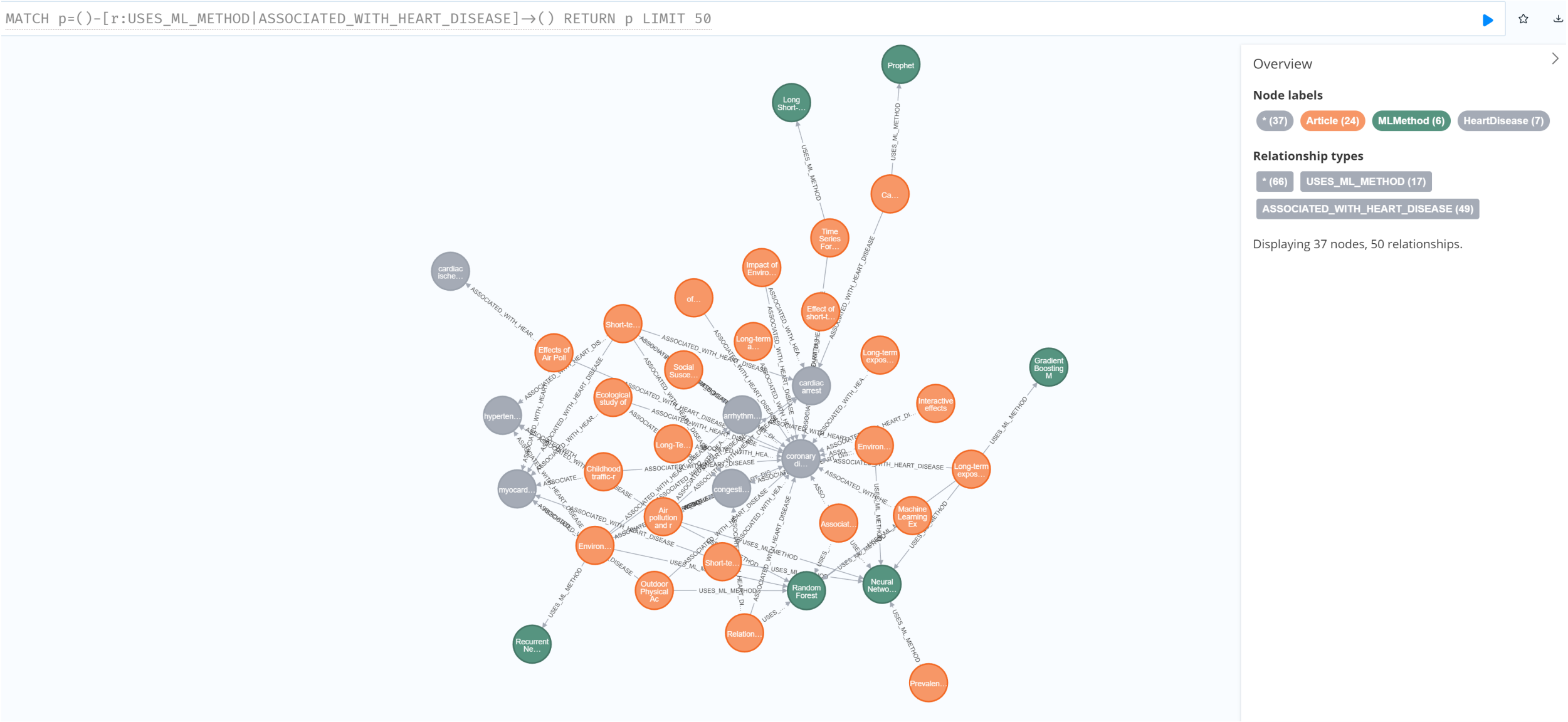

Figure 10: Query example

#### 3.9.8 Querying for Methodological Insights

This graph-based structure allows for analytical queries to uncover trends in the literature. For example, to identify the most frequently used machine learning methods within the subset of empirical studies examining ozone and heart disease, the following Cypher query is executed. This query traverses multiple relationship types to filter and aggregate the data, a task well-suited to a graph database.

```
MATCH (a:Article)
// Collect study type names per article
OPTIONAL MATCH (a)-[:STUDY_TYPE]->(st:StudyType)
WITH a, COLLECT(DISTINCT toLower(st.name)) AS studyTypeNames
// Filter for empirical studies only (exclude non-empirical types)
WHERE SIZE(studyTypeNames) > 0
  AND NONE(stName IN studyTypeNames WHERE stName IN [
    'review', 'systematic review', 'meta-analysis', 'expert opinion',
    'scoping review', 'dissertation/thesis', 'short communication',
    'methodological paper', 'theoretical study', 'report'
  ])
// Confirm articles explicitly have ozone, ML methods, and heart disease
AND EXISTS((a)-[:USES_ML_METHOD]->(:MLMethod))
AND EXISTS((a)-[:ASSOCIATED_WITH_HEART_DISEASE]->(:HeartDisease))
AND EXISTS {
  MATCH (a)-[:RELATED_TO_POLLUTANT]->(oz:PollutantTerm)
  WHERE toLower(oz.name) CONTAINS 'ozone'
}
// Exclude comment/reply-type articles explicitly
AND NOT toLower(a.title) CONTAINS 'comment'
AND NOT toLower(a.title) CONTAINS 'reply'
// Retrieve ML methods and count their occurrences
MATCH (a)-[:USES_ML_METHOD]->(ml:MLMethod)
RETURN
    ml.name AS `ML Method`,
    COUNT(DISTINCT a.doi) AS `Number of Articles`
ORDER BY `Number of Articles` DESC
```

| ML Method | Number of Articles |
|---|---|
| Random Forest | 20 |
| Gradient Boosting Machines | 14 |
| Support Vector Machines | 13 |
| Neural Networks | 12 |
| K-Means Clustering | 9 |
| Recurrent Neural Networks | 7 |
| Deep Neural Networks | 7 |
| Convolutional Neural Networks | 5 |
| Geographically Weighted Regression | 3 |
| Long Short-Term Memory Networks | 2 |
| Hierarchical Clustering | 2 |
| Markov Chain Monte Carlo | 2 |
| Autoencoders | 2 |
| Prophet | 2 |
| Principal Component Analysis | 1 |
| Extreme Learning Machines | 1 |
| Named Entity Recognition | 1 |

The execution of this query against the knowledge graph yields a clear distribution of methodologies employed across the 34 included studies. Tree-based ensemble methods are most prominent, with Random Forest being the predominant technique, utilized in 20 articles (58.8%). Gradient Boosting Machines and Support Vector Machines were also frequently applied, appearing in 14 and 13 studies, respectively. Neural network architectures showed considerable diversity, with the general category of Neural Networks reported in 12 studies, and more specific forms such as Recurrent and Deep Neural Networks each used in 7 studies. The data indicates a trend towards applying multiple ML techniques within a single study, with the 34 articles reporting 102 instances of ML method applications, averaging approximately three distinct methods per paper. This detailed, queryable insight into methodological trends is a result of structuring the extracted literature as a knowledge graph.

### 3.10 Stage 8: Vector Database Indexing

After structured field extraction and topic modeling, the pipeline creates vector-based search indices using Qdrant, a high-performance vector database. This stage transforms the processed documents into searchable embeddings that enable semantic retrieval for the downstream RAG (Retrieval-Augmented Generation) system. The indexing process is handled by the scripts/qdrant_indexing_standalone.py module, which operates independently of other pipeline components to ensure modularity.

#### 3.10.1 Dual Collection Architecture

The system builds two distinct Qdrant collections to serve different retrieval needs:

1. **PDF Chunks Collection**: Contains chunked full-text content from PDFs along with metadata. This collection enables retrieval of specific document sections for detailed analysis.
2. **Structured Data Collection**: Contains the structured fields extracted by the LLM along with core metadata. This collection enables retrieval based on specific research attributes like methodologies, pollutants, or study types.

Both collections use the same embedding model (text-embedding-3-large from OpenAI) to ensure semantic consistency between retrieval types. Collection names are prefixed with the active knowledge base identifier to support multiple research domains within the same Qdrant instance.

```
def create_or_recreate_collection(
    client: QdrantClient,
    collection_name: str,
    embed_dim: int = 3072,
    recreate: bool = False
) -> bool:
    """
    Create or recreate a Qdrant collection with specified parameters.
    """
    try:
        collections = client.get_collections().collections
        collection_exists = any(c.name == collection_name for c in collections)

        if collection_exists and recreate:
            logger.info(f"Deleting existing collection: {collection_name}")
            client.delete_collection(collection_name=collection_name)
        elif collection_exists:
            logger.info(f"Collection {collection_name} already exists, skipping creation")
            return True

        logger.info(f"Creating collection: {collection_name} with dimension {embed_dim}")
        client.create_collection(
            collection_name=collection_name,
            vectors_config=VectorParams(size=embed_dim, distance=Distance.COSINE)
        )
        return True

    except Exception as e:
        logger.error(f"Failed to create collection {collection_name}: {e}")
        return False
```

#### 3.10.2 Content Processing and Embedding

For the PDF chunks collection, the system concatenates core metadata (title, DOI, journal, citations) with the full text and table data extracted by Docling. This combined content is then split into overlapping chunks using tiktoken to respect the embedding model's context window (typically 7,000 tokens with 200-token overlap). Each chunk is processed through OpenAI's text-embedding-3-large model to generate 3,072-dimensional dense vectors.

```
def chunk_text(text: str, max_tokens: int = 7000, overlap: int = 200) -> List[str]:
    """
```

```
    Split text into overlapping chunks based on token count.
    """
    if not text or pd.isna(text):
        return []

    try:
        encoding = tiktoken.encoding_for_model("text-embedding-3-large")
        tokens = encoding.encode(str(text))

        if len(tokens) <= max_tokens:
            return [text]

        chunks = []
        start = 0

        while start < len(tokens):
            end = min(start + max_tokens, len(tokens))
            chunk_tokens = tokens[start:end]
            chunk_text = encoding.decode(chunk_tokens)
            chunks.append(chunk_text)

            if end >= len(tokens):
                break

            start += max_tokens - overlap

        return chunks

    except Exception as e:
        logger.warning(f"Error chunking text: {e}")
        return [text]
```

For the structured data collection, the system builds content strings from the metadata and structured fields extracted during the LLM stage. Fields are formatted as key-value pairs (e.g., "Tillage Practices: no-till, conservation tillage") to provide clear context for embedding generation. Since structured data is typically more concise, chunking is not required, with each document generating a single embedding.

### 3.10.3 Batch Processing and Storage

To handle hundreds of documents efficiently, the indexing process uses batch embedding generation and batch database insertion. The system processes texts in configurable batches (typically 25 texts per embedding request) to balance API efficiency with memory usage. Generated embeddings are accumulated and inserted into Qdrant in larger batches (typically 256 points) to minimize database overhead.

```
# Process chunks in embedding batches
for i in range(0, len(chunks), embed_batch_size):
    chunk_batch = chunks[i:i + embed_batch_size]
    embeddings = get_openai_batch_embedding(chunk_batch, embed_model, api_key)

    if not embeddings:
        continue

    # Create points for this batch
    for chunk_idx, (chunk, embedding) in enumerate(zip(chunk_batch, embeddings)):
        payload = {
            "doc_idx": int(idx),
            "chunk_idx": chunk_idx + i,
            "content": chunk,
            "qdrant_point_id": idx * 10000 + chunk_idx + i
        }
```

```
            # Add metadata
            for col in core_metadata:
                if col in row and pd.notna(row[col]):
                    payload[col] = str(row[col])

            point_id = idx * 10000 + chunk_idx + i
            point = rest_models.PointStruct(
                id=point_id,
                vector=embedding,
                payload=payload
            )
            points_batch.append(point)

        # Upsert when batch is full
        if len(points_batch) >= batch_size:
            client.upsert(
                collection_name=collection_name,
                points=points_batch,
                wait=True
            )
            points_batch = []
```

Each stored point includes the embedding vector, the original text content, and a comprehensive payload containing document metadata, chunk indices, and extracted fields. Numeric point IDs are generated deterministically based on document and chunk indices to ensure consistency across pipeline runs and enable point updates if needed.

#### 3.10.4 Integration with Knowledge Base Architecture

The indexing stage integrates with the broader knowledge base architecture by using the selected knowledge base configuration to determine collection naming and connection parameters. Collection names incorporate the knowledge base prefix (e.g., "wastewater_surveillance_pdf_chunks") to maintain separation between different research domains. The stage verifies successful collection creation and reports the final document and chunk counts. Upon completion, the Qdrant collections are ready to serve the RAG system, enabling semantic search across both full-text content and structured research attributes to support comprehensive literature analysis and synthesis.

### 3.11 ETL Pipeline Summary and Output

The sequential execution of these eight stages transforms a collection of raw, unstructured PDF documents into a comprehensive, semantically unified, and queryable research infrastructure. The process is designed to be robust and efficient, using parallel processing for I/O-bound tasks like API calls and GPU-bound tasks like document parsing. Configurable heuristics are employed at multiple stages, from pre-emptive artifact filtering and layout post-processing to semantic unification, to ensure the quality and consistency of the final data product.

The pipeline produces two complementary data products that together form the foundation for downstream analysis and retrieval systems:

**Neo4j Knowledge Graph**: A graph database where each :Article node represents a rich, interconnected entity linked through explicit relationships to authors, key concepts, methodologies, findings, and thematic context within the broader literature. This structured representation enables complex multi-faceted queries that traverse relationships between entities, such as identifying methodological trends across specific research domains or discovering knowledge gaps in particular topic areas.

**Qdrant Vector Collections**: Two specialized vector databases that enable semantic search capabilities. The PDF chunks collection provides fine-grained retrieval of document sections, while the structured data collection enables search based on extracted research attributes. These collections use high-dimensional embeddings to capture semantic meaning, allowing the system to retrieve relevant content based on conceptual similarity rather than exact keyword matching.

This dual architecture combines the precision of structured queries with the flexibility of semantic search. The knowledge graph excels at relationship-based analysis and categorical filtering, while the vector collections enable natural language queries and conceptual exploration. Together, they serve as the foundational "non-parametric memory" for the Retrieval-Augmented Generation (RAG) system, enabling nuanced, context-aware information synthesis to address complex research questions with high fidelity and complete traceability back to source documents.

### 3.12 HySemRAG-QA Framework: An Agentic Approach for Verifiable Generation

With the creation of a structured knowledge graph, the pipeline transitions from data ingestion to its primary purpose of enabling users to ask complex questions and receive accurate, synthesized, and verifiable answers. This is accomplished through the Hybrid Semantic Retrieval-Augmented Generation with Quality Assurance (HySemRAG-QA) framework, a system designed to address the limitations of standard RAG architectures, namely noisy retrieval and LLM hallucination. The system's architecture uses a multi-layered approach to ensure the reliability of its final output, from initial query to final validation.

The core of the framework creates a chain of custody for every piece of information. This begins with a hybrid retrieval engine that goes beyond simple vector search. It combines results from three distinct sources: semantic search (Qdrant), keyword search, and structured graph traversals (Neo4j), and uses Reciprocal Rank Fusion (RRF) to produce a single, relevant context.

This context is then passed to an agentic self-correction framework. This is not a single-pass generation process. It implements a multi-agent system where a primary "generator" LLM drafts a cited answer, and a secondary "QA agent" LLM audits that output for factual accuracy, logical consistency, and adherence to citation protocols. If the QA agent detects a flaw, it provides corrective feedback, and the generator revises its work in an iterative loop until a reasoned answer is produced.

Finally, to ensure verifiability, the system performs a post-hoc audit. Every citation in the final, agent-approved answer is checked against the ground-truth database to confirm its provenance. As a byproduct of this multi-stage validation process, the system generates a comprehensive interaction log, creating a preference dataset of (prompt, rejected_answer, chosen_answer) triplets suitable for AI model training and fine tuning. The following sections will detail the technical implementation of each of these modules.

#### 3.12.1 Hybrid Retrieval with Reciprocal Rank Fusion

The framework employs a three-source retrieval strategy that combines semantic search, keyword search, and knowledge graph traversal. Each query is processed through:

- **Semantic Search**: Vector similarity using OpenAI's text-embedding-3-large model (3,072 dimensions) against Qdrant collections
- **Keyword Search**: Full-text filtering using Qdrant's MatchText functionality
- **Knowledge Graph Search**: Entity extraction and Neo4j traversal using dynamically loaded unification functions

Results from all three sources are merged using Reciprocal Rank Fusion (RRF) with the formula:

$$\text{rrf_score} = \sum_{source} \frac{1}{K + rank_{source}}$$

where K=60 and ranks beyond the top-k threshold contribute zero to the final score.

#### 3.12.2 Dynamic Knowledge Base Routing

The system supports multiple isolated knowledge bases through a dynamic configuration manager. Each knowledge base maintains:

- **Isolated Data Sources**: Separate Neo4j databases, Qdrant collection prefixes, and ETL configurations
- **KB-Specific Entity Extraction**: Dynamic loading of unification functions based on selected extraction columns
- **Automatic User Assignment**: Research area keyword matching to route users to appropriate knowledge bases
- **Environment-Aware Deployment**: Automatic path translation between host and container environments

This architecture enables the same RAG system to serve multiple research domains simultaneously while maintaining complete data isolation.

#### 3.12.3 Iterative Chain-of-Thought with Quality Assurance

The core reasoning process implements a multi-agent validation loop:

1. **Generator Agent**: Primary LLM (Claude Sonnet 4) drafts structured responses with explicit metadata requirements
2. **Evaluator Agent**: Secondary LLM (Gemini 2.5 Flash) audits outputs for accuracy and citation compliance
3. **Iterative Refinement**: Up to 3 regeneration cycles based on evaluator feedback
4. **Final QA Check**: Independent validation against source data with strict citation requirements

Each observation must include precise metadata following a non-negotiable schema: - PDF sources: PDF_DocIndex, PDF_ChunkIndex, DOI, InTextCitation, FullCitation, ZoteroKey - Structured sources: Struct_DocIndex, Struct_ChunkIndex, plus citation metadata - Knowledge Graph sources: KG_DocIndex, Relation, plus citation metadata

The system enforces single-source observations and rejects mixed-source citations to prevent hallucination.

#### 3.12.4 Dynamic Query Enhancement

User queries undergo enhancement through:

1. **LLM-Based Reformulation**: Claude Sonnet 4 rewrites queries in standard domain language and extracts 3-5 content keywords
2. **Dynamic Entity Extraction**: KB-specific unification functions map query terms to knowledge graph entities
3. **Contextual Enrichment**: Neo4j queries provide additional context based on extracted entities

Entity extraction uses dynamically loaded functions based on each KB's field mappings, ensuring the system adapts to different research domains without manual reconfiguration.

#### 3.12.5 Post-Hoc Citation Verification

Every generated response undergoes verification against ground-truth data:

- **Metadata Validation**: DOI and ZoteroKey matching against the canonical ETL DataFrame
- **Index Verification**: Doc_idx and chunk_idx validation against retrieved chunks
- **Content Similarity**: Levenshtein distance comparison between cited evidence and source text
- **Validity Classification**: Automatic labeling as VALID ($\geq$ 0.8), POSSIBLY_VALID (0.5-0.8), or INVALID (<0.5)

This creates a comprehensive audit trail and enables continuous quality monitoring of the RAG system's factual accuracy.

#### 3.12.6 KB-Aware Session Tracking

The system maintains detailed interaction logs that serve dual purposes:

1. **User Session Management**: KB-specific session folders track all interactions, retrieval results, and reasoning chains
2. **Training Data Generation**: Each interaction produces (query, rejected_answer, accepted_answer) triplets suitable for preference learning

Sessions include complete provenance information: retrieval sources, agentic reasoning steps, QA verdicts, and similarity scores, creating a rich dataset for future model improvement.

## 4 Results

### 4.1 System Performance Evaluation

We evaluated HySemRAG using 643 observations across 60 testing sessions conducted during system development. While this represents controlled developer testing rather than independent evaluation, it provides insight into system capabilities and limitations.

**Retrieval Performance**: Structured field extraction achieved higher semantic similarity scores (0.655 $\pm$ 0.178, n=165) compared to PDF text chunking (0.485 $\pm$ 0.204, n=263). Mann-Whitney U testing confirmed this 35.1% performance difference as statistically significant ($p < 0.000001$).

**Evidence Quality Distribution**: Classification revealed 100 observations (24.5%) as high quality ($\geq$ 70% similarity), 258 observations (63.2%) as medium quality (30-69% similarity), and 50 observations (12.3%) as low quality (<30% similarity).

**Quality Assurance Mechanisms**: The agentic self-correction system demonstrated 68.3% single-pass success rate, requiring average 2.33 iterations per query. Quality-controlled analysis of 394 verified observations (61.3% acceptance rate) showed citation accuracy of 99.0% $\pm$ 4.5% and semantic similarity scores of 1.000 $\pm$ 0.000.

### 4.2 Citation Accuracy and Verification

Citation component analysis revealed field-level accuracy: DOI matching (74.4%), in-text citations (73.9%), full citation formatting (73.9%), and Zotero key matching (72.3%). Vector database integration achieved 99.6% success rate (242 of 243 attempts), indicating stable retrieval performance.

### 4.3 Temporal Stability and Error Analysis

The system exhibited temporal instability with performance declining at -1.9% per day over the evaluation period. Daily performance ranged from 100.0% valid responses at peak to 25.0% at worst. Session-level analysis showed 27 sessions (45.0%) achieving $\geq$ 80% valid responses while 20 sessions (33.3%) produced <50% valid responses.

Complete response parsing was achieved for 518 of 643 observations (80.6%), with 125 parsing failures (19.4%) including 65 empty metadata errors and 55 malformed structure errors.

### 4.4 Domain-Specific Application Results

Applied to geospatial epidemiology literature on ozone and cardiovascular disease, the system identified methodological trends across 34 empirical studies. Tree-based ensemble methods dominated, with Random Forest utilized in 20 articles (58.8%), Gradient Boosting Machines in 14 studies, and Support Vector Machines in 13 studies. Neural network architectures showed diversity, with general Neural Networks reported in 12 studies and specialized forms like Recurrent and Deep Neural Networks each used in 7 studies.

Knowledge graph queries revealed that 34 articles reported 102 instances of ML method applications, averaging approximately three distinct methods per paper, indicating trends toward multi-method approaches within individual studies.

## 5 Discussion

### 5.1 Performance and Limitations

The 61.3% acceptance rate meets minimum thresholds for research applications but requires improvement for reliable deployment. The advantage of structured field extraction over PDF chunking (35.1% improvement) suggests semantic specialization preserves document organization and aligns better with natural query patterns targeting methodological concepts.

Temporal instability (-1.9% daily decline) and session heterogeneity indicate system sensitivity to operational conditions not representative of production environments. This systematic degradation suggests model drift or prompt degradation rather than stable performance, requiring investigation before deployment.

### 5.2 Methodological Contributions

The semantic unification pipeline represents a contribution, using sentence transformers with cosine similarity and threshold-based mapping to resolve terminological ambiguity with broad applicability. The agentic self-correction framework addresses RAG limitations, though modest improvement from iterative refinement suggests the approach needs refinement.

Custom Docling modifications demonstrate practical solutions to real-world PDF processing challenges, particularly for scholarly documents with complex layouts. The hybrid retrieval approach combining semantic search, keyword filtering, and knowledge graph traversal provides comprehensive coverage addressing individual method limitations.

### 5.3 Broader Implications

The framework's modular design enables application across research domains beyond geospatial epidemiology. Knowledge graph construction facilitates complex analytical queries impossible with traditional relational approaches, while vector collections enable semantic search capturing conceptual similarity beyond keyword matching.

The system's ability to identify methodological trends and gaps demonstrates potential for accelerating research discovery and informing funding priorities. Complete citation verification and provenance tracking address reproducibility concerns in AI-generated research synthesis.

## 6 Conclusion

We present HySemRAG, a framework addressing limitations in automated literature synthesis through integrated ETL processing and multi-agent RAG capabilities. The system demonstrates improvements in retrieval accuracy and citation verification while enabling systematic identification of methodological gaps across research domains.

Key contributions include: (1) hybrid retrieval strategies combining multiple search modalities with Reciprocal Rank Fusion, (2) document layout analysis modifications addressing scholarly PDF processing challenges, (3) semantic unification resolving terminological heterogeneity, (4) multi-agent quality assurance with iterative refinement, and (5) comprehensive citation verification ensuring complete provenance tracking.

Evaluation reveals structured field extraction achieving 35.1% higher semantic similarity compared to PDF chunking approaches, with the system identifying methodological trends in geospatial epidemiology literature. While temporal instability and session heterogeneity indicate areas for improvement, the framework provides a foundation for accelerating evidence synthesis and research discovery.

Future work should address system stability through independent user studies, cross-domain evaluation, and longitudinal performance assessment under real-world conditions. The modular architecture enables extensions to diverse scientific domains, potentially transforming how researchers discover and synthesize evidence across disciplinary boundaries.